\documentclass[preprint,12pt]{elsarticle}

\usepackage{amssymb}
\usepackage{amsmath}

\usepackage{ulem}
\usepackage{epsfig}
\usepackage{graphicx}
\usepackage{subfigure}
\usepackage{booktabs}
\usepackage{color}
\usepackage{float}
\usepackage{natbib}
\usepackage{tikz}
\usetikzlibrary{shapes,arrows}
\usetikzlibrary{matrix,arrows,positioning}
\usetikzlibrary{decorations.pathreplacing, calligraphy}
\usepackage{longtable}
\usepackage{multirow}
\usepackage{hyperref}
\usepackage{verbatim,caption}
\usepackage{listings}
\usepackage{pdfpages}
\usepackage{tabularray}

\newtheorem{pro}{Proposition}[section]

\newtheorem{defn}{Definition}[section]

\newtheorem{result}[pro]{Result}

\usepackage{lineno}

\journal{}

\begin{document}

\begin{frontmatter}


 \cortext[cor1]{}
 
\title{A Tutorial on Structural Identifiability of Epidemic Models Using \textit{StructuralIdentifiability.jl}} 


\author[1]{Yuganthi R. Liyanage\corref{cor1}} 
\ead{aliyanage2018@fau.edu}
\author[2]{Omar Saucedo} 
\author[1]{Necibe Tuncer } 
\author[3,4]{Gerardo Chowell\corref{cor1}} 
\ead{gchowell@gsu.edu}

\affiliation[1]{organization={Department of Mathematics and Statistics, Florida Atlantic University},
            addressline={}, 
            city={Boca Raton},
            postcode={33431}, 
            state={Florida},
            country={USA}}

\affiliation[2]{organization={Department of Mathematics, Virginia Tech},
            addressline={}, 
            city={Blacksburg},
            postcode={24060}, 
            state={Virginia},
            country={USA}}    

\affiliation[3]{organization={School of Public Health, Georgia State University},
            addressline={}, 
            city={Atlanta},
            postcode={30303}, 
            state={Georgia},
            country={USA}}   

\affiliation[4]{organization={Department of Applied Mathematics,  Kyung Hee University},
            addressline={}, 
            city={Yongin},
            postcode={17104}, 
            state={},
            country={Korea}}

\begin{abstract}

Structural identifiability is the theoretical ability to uniquely recover model parameters from ideal, noise-free data and is a prerequisite for reliable parameter estimation in epidemic modeling. Despite its relevance in model calibration and inference, structural identifiability analysis remains underused and inconsistently applied in the infectious disease modeling literature. This paper presents a user-oriented methodological tutorial that demonstrates how global structural identifiability analysis can be systematically integrated into epidemic modeling workflows. We provide a reproducible workflow for conducting structural identifiability analysis of ordinary differential equation models using the Julia package \textit{StructuralIdentifiability.jl}. We illustrate this workflow across a range of commonly used epidemic models, including SEIR variants with asymptomatic and presymptomatic transmission, vector-borne disease models, and systems incorporating hospitalization and disease-induced mortality. In addition to hands-on methodological guidance, we introduce a novel visual communication strategy that embeds identifiability results directly into compartmental diagrams, facilitating interpretation and interdisciplinary communication. Our results illustrate, through a unified and reproducible workflow, how identifiability depends critically on model structure, the choice of observed variables, and assumptions about initial conditions, and how identifiable parameter combinations can be recovered even when individual parameters are not globally identifiable. Rather than introducing new identifiability theory, this work focuses on the practical implementation, interpretation, and communication of existing methods in applied epidemic modeling contexts. By combining practical instruction, comparative insights across model classes, and enhanced visualization tools, this work serves as both a reference and a teaching resource for researchers and educators seeking to incorporate structural identifiability analysis into epidemic model design and interpretation. All code and annotated diagrams are publicly available to support reproducibility and reuse.

\end{abstract}



\end{frontmatter}


\section{Introduction}
Differential equation-based mathematical models provide a rigorous and versatile quantitative framework to explore the dynamics of complex systems across disciplines such as medicine, epidemiology, and biology. These models consist of systems of differential equations defined by initial conditions and parameters that govern the temporal evolution of state variables. In epidemiology, they are widely used to quantify transmission dynamics, estimate parameters such as reproduction numbers and infectious periods, assess the impact of interventions, and support public health decision-making \cite{brauer2008, arino2007, chowell2004, anderson2004, saucedo2019computing}. Beyond simulating disease progression, these models are essential for hypothesis testing and mechanistic inference. However, the validity of such inferences depends on whether parameters can be uniquely estimated from the available data, which is often limited to case counts, hospitalizations, or deaths \cite{yan2019, banks2009}. This requirement makes structural identifiability analysis a crucial first step: it determines, under ideal noise-free conditions, whether unique parameter estimates are theoretically possible~\cite{tuncer2018, roosa2019}. As such, structural identifiability constitutes a necessary theoretical prerequisite for parameter estimation and complements, rather than replaces, local or sensitivity-based analyses. Neglecting this step risks parameter non-identifiability, leading to unreliable estimates and potentially misguided policy recommendations ~\cite{villaverde2016, eisenberg2013,massonis2021structural,dankwa2022structural,gallo2022lack}.\\

\begin{figure}[H]
    \centering
\includegraphics[scale=0.15]{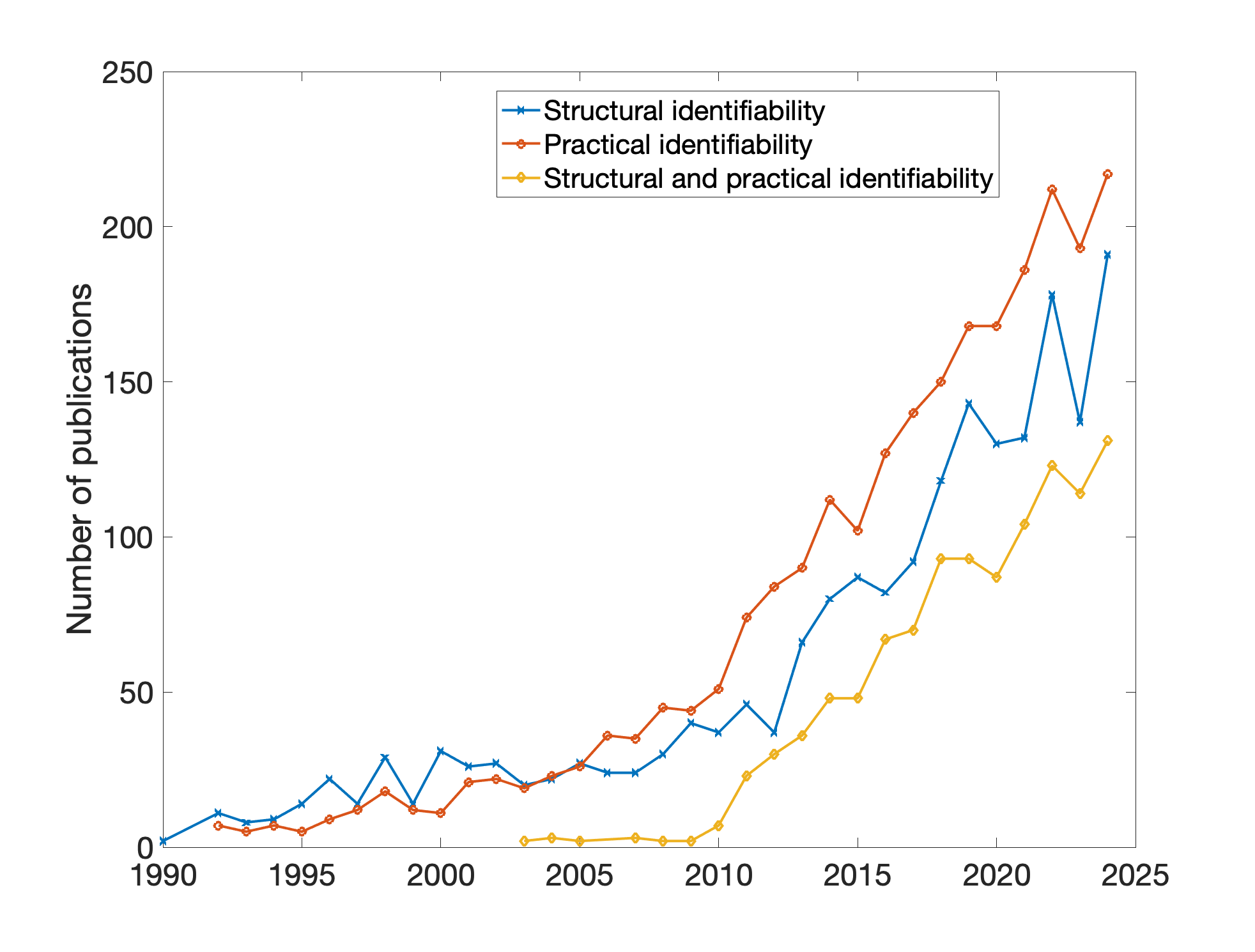}
    \caption{As illustrated in this figure, the number of publications that include structural identifiability, practical identifiability, or both in their titles has increased markedly between 1990 and 2024. This trend reflects the growing recognition within the modeling community of the importance of rigorous identifiability analyses to support reliable parameter inference. Notably, practical identifiability has received the most attention in recent years, while interest in studies that integrate both structural and practical perspectives continues to rise. Data retrieved from the Web of Science core collection.}
    \label{fig:NumberofPublications}
\end{figure}

Over the past decade, structural identifiability analysis of ordinary differential equation–based models has reached methodological maturity, supported by a diverse ecosystem of symbolic and numerical tools and clarified through recent benchmarking and synthesis studies ~\cite{heinrich2025structural}. Despite these advances, identifiability analysis remains underused and inconsistently addressed in the infectious disease modeling literature (Figure \ref{fig:NumberofPublications}), particularly in partially observed systems where parameter confounding and non-identifiable parameter combinations are common. Thus, current challenges lie less in the development of new identifiability theory and more in the effective application, interpretation, and communication of identifiability results within applied modeling workflows.

Recent benchmarking and review studies have provided comprehensive overviews of available structural identifiability methods and software tools, clarifying their respective strengths, limitations, and methodological trade-offs ~\cite{rey2023benchmarking, heinrich2025structural}. The present study does not aim to replicate or extend these theoretical or benchmarking efforts. Instead, it adopts a user-oriented perspective focused on applied epidemic modeling, illustrating how global structural identifiability analysis can be systematically integrated into model development, interpretation, and communication.

We previously introduced a workflow for structural identifiability analysis using DAISY ~\cite{bellu2007, Chowell2023}. While DAISY represented an important advance, its limitations for larger or more complex models motivate the tutorial focus on the Julia package StructuralIdentifiability.jl ~\cite{structidjl2023}, which provides an efficient implementation of differential algebra–based methods for identifiability assessment. In this work, we build on that foundation by presenting a reproducible and user-oriented tutorial workflow based on StructuralIdentifiability.jl, illustrating how global structural identifiability analysis can be integrated into applied epidemic modeling workflows.

Our contribution is primarily methodological and pedagogical. First, we provide a practical tutorial showing how to use \textit{StructuralIdentifiability.jl} to evaluate identifiability in compartmental epidemic models, addressing gaps in prior work where structural identifiability analysis has not always been systematically verified in epidemiological modeling studies. Second, we propose a novel way to communicate identifiability by embedding results directly into compartmental diagrams. By visually annotating whether parameters such as transmission ($\beta$) or recovery ($\gamma$) rates are globally identifiable, non-identifiable, or conditionally identifiable, researchers and decision-makers gain immediate insight into the reliability of parameter inference. Unlike standard compartmental diagrams that depict only model structure, our approach encodes identifiability directly within the diagram, enabling rapid detection of parameter confounding and non-identifiable structures, and supporting clearer model interpretation. This approach is intended to enhance transparency, support interdisciplinary communication, and serve as a training tool for students and practitioners.\\

Structural identifiability is a theoretical property that precedes practical concerns such as noisy or sparse data and provides a foundation for determining whether parameter estimation is even possible. Figure \ref{fig:NumberofPublications} illustrates the growth of publications addressing identifiability since 1990, reflecting increasing recognition of its importance and the availability of computational tools that facilitate its application. Several methods exist for structural identifiability analysis—including Taylor series~\cite{pohjanpalo1978}, generating series~\cite{walter1982}, similarity transformations~\cite{vajda1989}, and direct tests~\cite{joly1998some}—but the differential algebra approach remains the most widely used, particularly for epidemic models~\cite{ljung1994, bellu2007}. In recent years, comprehensive benchmarking and review studies have provided a broader perspective on the landscape of structural identifiability methods and software tools, highlighting their relative strengths, computational trade-offs, and domains of applicability (e.g.,\cite{heinrich2025structural,rey2023benchmarking}). These studies emphasize the growing maturity of the field and the importance of selecting appropriate methods based on model complexity and data structure. In this work, we focus on demonstrating how such established methods can be applied in practice rather than introducing new theoretical developments. Using StructuralIdentifiability.jl, we demonstrate how this method can be efficiently implemented in practice and applied to increasingly complex epidemic modeling settings.\\

Related work by the authors has examined structural and practical identifiability in the context of phenomenological growth models for epidemic forecasting ~\cite{phenom2025}. The present study complements that work by focusing on mechanistic compartmental epidemic models and on global structural identifiability properties, while explicitly adopting a user-oriented tutorial perspective rather than introducing new structural identifiability theory or performing software benchmarking.

Overall, this work aims to bridge the gap between theoretical identifiability methods and their practical use in epidemic modeling by providing a reproducible workflow, comparative insights across model structures, and tools to facilitate interpretation and communication of identifiability results.\\
 
Structural identifiability is an intrinsic property of a model, governed by the identifiability of its individual parameters. Fundamentally, structural identifiability analysis asserts that two distinct sets of parameter values can yield the same model output only if they are identical. Over the years, several methods have been developed to assess structural identifiability, including classical approaches and modern differential algebra-based methods~\cite{villaverde2016, bellu2007, pohjanpalo1978,walter1982,vajda1989,joly1998some,ljung1994,miao2011identifiability, Saucedo2024Indent,ligon2018genssi, diaz2023strike,anstett2020priori, castro2020testing, saccomani2003parameter}.\\

The structural identifiability of a model is based on the fact that observations vary as the model parameters vary. Suppose that another set of parameters, denoted by $\hat{\mathbf{p}}$, produces the same observations as those denoted by $y_1(t, \mathbf{p})$. That is,
\[
y_1(t, \mathbf{p}) = y_1(t, \hat{\mathbf{p}}).
\]
By definition, this should only happen when the two parameter sets are identical if the model parameters are structurally identifiable ~\cite{guillaume2019introductory,cobelli1980parameter,bellman1970structural,distefano1980parameter}. That is, $\mathbf{p} = \hat{\mathbf{p}}$. Hence, we define the structural identifiability of the model parameters as follows:

\begin{defn}\label{def1}
Let $\mathbf{p}$ and $\hat{\mathbf{p}}$ be distinct model parameters, and let $y_1(t, \mathbf{p})$ be the observations. If
\[
y_1(t,\mathbf{p}) = y_1(t, \hat{\mathbf{p}}) \quad \text{implies} \quad \mathbf{p} = \hat{\mathbf{p}},
\]
then we conclude that the model is structurally identifiable from noise-free and continuous observations $y_1(t)$.
\end{defn}

This definition is equivalent to the differential algebra formulation used in \texttt{StructuralIdentifiability.jl}, where identifiability is assessed by determining whether parameters can be uniquely recovered from the observable outputs and their derivatives.

\textbf{Overview of the algorithm behind StructuralIdentifiability.jl}

The algorithm implemented in the Julia package \texttt{StructuralIdentifiability.jl}, as introduced in \cite{structidjl2023}, is based on the classical input-output approach to structural identifiability. This method was originally proposed by Ollivier in \cite{Ollivier1990} and builds upon the classical input–output differential algebra framework \cite{ljung1994}.

To illustrate the approach, consider the following system of differential equations with one output variable:

\begin{equation}\label{eq:example_system}
\begin{cases}
x_1' = a x_2, \\
x_2' = -b x_1, \\
y = x_1 + c,
\end{cases}
\end{equation}

The core idea of the input-output method is to eliminate unobserved state variables, thereby deriving equations that involve only model parameters, inputs, and observable outputs (and their derivatives). Ideally, a minimal set of such equations is derived. Classical approaches require these to form a \textit{characteristic set} \cite{Ollivier1990}, but \texttt{StructuralIdentifiability.jl} uses a slightly different notion of minimality tailored for symbolic computation \cite{structidjl2023}. \\

For the system above, the algorithm yields the following minimal relation between the parameters and the output \( y \):

\[
y'' + ab y - abc = 0.
\]

This equation indicates that the observable output \( y(t) \) depends only on the combinations \( ab \) and \( abc \). From an identifiability perspective, this means we cannot recover parameters \( a, b, c \) individually from \( y(t) \), but only the combinations \( ab \) and \( abc \). \\ 

If we evaluate this equation at two time points \( t_1 \) and \( t_2 \), we can formulate the following linear system:

\[
\begin{bmatrix}
y(t_1) & 1 \\
y(t_2) & 1
\end{bmatrix}
\begin{bmatrix}
ab \\
-abc
\end{bmatrix}
=
- \begin{bmatrix}
y''(t_1) \\
y''(t_2)
\end{bmatrix}.
\]

If the matrix on the left-hand side is nonsingular, we can uniquely recover \(ab\) and 
\(abc\). This follows from standard linear algebra: nonsingularity ensures that the system admits a unique solution, which implies injectivity of the mapping from parameter combinations to observable quantities and hence structural identifiability in the sense of Definition \ref{def1}\cite{miao2011identifiability}.
Consequently, the parameter 
\(c=\displaystyle\frac{abc}{ab}\) is identifiable, while \(a\) and \(b\) are not individually identifiable.


In essence, structural identifiability reduces to a \textit{field membership problem}: determining whether each parameter lies in the differential field generated by the observed outputs and their derivatives. \\ 

Importantly, the validity of this method relies on the nonsingularity of certain matrices arising from evaluations of the observed outputs and their derivatives at finitely many time points, as illustrated in Example \ref{eq:example_system}. These matrices represent linear maps from identifiable parameter combinations to observed data. Nonsingularity ensures that these maps are injective and thus that the corresponding parameter combinations can be uniquely recovered from the observations.

While most software packages assume nonsingularity by default, this assumption may fail in practice. As discussed in \cite{Hong2020} (Example 2.14), such failures correspond to a loss of injectivity and can lead to incorrect conclusions about identifiability if not detected. \texttt{StructuralIdentifiability.jl} addresses this issue by checking for matrix nonsingularity and issuing a warning when the assumption is violated.


\section{Overview of the StructuralIdentifiability.jl Toolbox}

The \texttt{StructuralIdentifiability.jl} package is a Julia-based symbolic computation toolbox designed to determine whether the parameters of a symbolic ordinary differential equation (ODE) model can be uniquely recovered from perfect, noise-free data \cite{structidjl2023}. It provides an efficient implementation of the input–output differential algebra approach, enabling researchers to assess global, local (unique up to discrete symmetries), or non-identifiability of model parameters directly from model equations without numerical simulation. While the package implements advanced symbolic algorithms, our focus here is not on the internal computational details but on its practical use within epidemic modeling workflows; we refer interested readers to ref. \cite{structidjl2023} for a comprehensive description of the underlying methods.

The core workflow involves three main steps:

\begin{enumerate}
    \item \textbf{Model specification:} Define the system of differential equations and observable outputs using the macro \texttt{@ODEmodel}.
    \item \textbf{Identifiability assessment:} Evaluate whether each parameter and initial condition is structurally identifiable using \texttt{assess\_identifiability(ode)}.
    \item \textbf{Identification of parameter combinations:} Use \texttt{find\_identifiable\_functions(ode)} to identify algebraic combinations of parameters that are structurally identifiable even when individual parameters are not.
\end{enumerate}

These steps provide a practical and reproducible framework to examine how model structure, observability, and assumptions about initial conditions influence identifiability outcomes.

A typical analysis begins by defining the model structure and observables. Below is an illustrative example demonstrating how to set up and analyze a simple SIR model:

\begin{verbatim}
using StructuralIdentifiability

# Define a simple SIR model with infection and recovery dynamics
ode = @ODEmodel(
    S'(t) = -beta*S(t)*I(t)/N,
    I'(t) = beta*S(t)*I(t)/N - gamma*I(t),
    R'(t) = gamma*I(t),
    y(t)  = I(t)     # Observable output: infectious population
)

# Assess parameter identifiability
assess_identifiability(ode)

# Identify combinations of parameters that are jointly identifiable
find_identifiable_functions(ode)
\end{verbatim}

The output of \texttt{assess\_identifiability()} classifies each parameter as \texttt{:globally}, \texttt{:locally}, or \texttt{:nonidentifiable}. A result of \texttt{:globally} means that the parameter can be uniquely recovered under all conditions, \texttt{:locally} indicates uniqueness up to discrete symmetries, and \texttt{:nonidentifiable} implies that multiple parameter values can produce identical outputs. The function \texttt{find\_identifiable\_functions()} lists algebraic combinations (e.g., ratios or products such as \(\beta/N\)) that are identifiable even when individual parameters are not. Users can optionally specify known initial conditions using the argument \texttt{known\_ic = [S, I, R]}, which may alter identifiability results.

The package supports systems with multiple observables, user-defined constants, and symbolic initial conditions. Compared to classical tools such as DAISY \cite{bellu2007}, \textit{StructuralIdentifiability.jl} offers efficient symbolic computations, automatically checks algebraic assumptions like matrix nonsingularity, and has been successfully applied to moderately high-dimensional epidemic models commonly used in practice. The emphasis here is on transparency, reproducibility, and interpretability for epidemic modeling applications.

The tutorial examples that follow illustrate this workflow across a range of compartmental epidemic models, demonstrating how identifiability depends on model structure, observability, and initial condition assumptions. The goal is to provide applied guidance on implementation, interpretation, and communication of identifiability results in epidemiological settings. All code and annotated notebooks used in this tutorial are available at the project’s \href{https://github.com/YuganthiLiyanage/Structural-identifiability-of-epidemic-models}{GitHub repository}.

\section{Structural Identifiability of epidemic
models}

In this section, we apply the identifiability workflow described in the previous section to a sequence of commonly used epidemic model structures. The goal is not to introduce new models or new theoretical identifiability results, but to illustrate how global structural identifiability analysis can be systematically conducted, interpreted, and communicated across representative epidemic modeling scenarios. Each example highlights how model structure, observable quantities, and assumptions about initial conditions shape identifiability outcomes, revealing recurring patterns and practical implications relevant to epidemic model design. We note that all identifiability results presented in this section are obtained using symbolic computation tools (primarily StructuralIdentifiability.jl) and are summarized to facilitate interpretation rather than to establish new theoretical results.

To enhance transparency and contextualize the results obtained using \textit{StructuralIdentifiability.jl}, we additionally evaluated selected models using two widely used structural identifiability tools, SIAN and STRIKE-GOLDD. The resulting identifiability classifications are reported alongside those obtained with StructuralIdentifiability.jl in the corresponding tables. This comparison is intended to illustrate how different symbolic approaches may yield consistent or divergent conclusions depending on model structure, observables, and underlying algorithmic assumptions, rather than to benchmark software performance or suggest methodological superiority.

\subsection{SEIR model}

The SEIR model is a foundational compartmental framework widely used to describe the dynamics of infectious diseases. It stratifies the population into four compartments: susceptible individuals ($S(t)$), exposed (latent) individuals ($E(t)$), infectious individuals ($I(t)$), and recovered individuals ($R(t)$). Infection occurs through a standard incidence term, $\beta\frac{S(t)I(t)}{N}$, where $I(t)/N$ captures the probability of contact between susceptible and infectious individuals. Exposed individuals transition to the infectious stage at rate $k$, and infectious individuals recover at rate $\gamma$. The model assumes a closed population, keeping the total population size $N$ constant.\\

Structural identifiability refers to the theoretical ability to recover model parameters uniquely from perfect noise-free observations of the system output. To examine the structural identifiability of the SEIR model parameters, we consider the observable output to be the number of new infections per unit of time, given by $y_1(t) = k E(t)$. Our goal is to determine whether the transmission rate $\beta$, the transition rate $k$, and the recovery rate $\gamma$ can be uniquely inferred from this observation.

The SEIR model is given by the following system of ordinary differential equations:
\begin{equation}\label{Model1}\tag{M$_1$}
\textbf{Model 1:}
\begin{cases}
\displaystyle\frac{dS}{dt} = -\beta\frac{SI}{N}, \quad S(0)=S_0\\[1.5ex]
\displaystyle \frac{dE}{dt} = \beta\frac{SI}{N} - kE,\quad E(0)=E_0  \\[1.5ex]
\displaystyle \frac{dI}{dt} = kE-\gamma I, \quad I(0)=I_0 \\[1.5ex]
\displaystyle \frac{dR}{dt} = \gamma I, \quad R(0)=R_0.
\end{cases}
\end{equation}

We used the \texttt{\texttt{StructuralIdentifiability.jl}} package in Julia to perform the identifiability analysis of the SEIR model. The analysis was carried out under two conditions: unknown initial conditions and known initial conditions.

The corresponding Julia input and output for both scenarios—assuming either unknown or known initial conditions—are shown below to demonstrate the structural identifiability analysis workflow. When all initial conditions are known, the total population size $N$ is also determined and must be treated as a known quantity in the analysis. As a result, $N$ is incorporated as an additional observable, which enables the identifiability of the transmission rate $\beta$. This distinction is reflected in Table \ref{SI_M1}\textbf{B}, where known initial conditions lead to full parameter identifiability.

In Julia, the model is specified using \texttt{@ODEmodel}. The following code block sets up the SEIR model equations, specifies the observable, and prepares the model for structural identifiability analysis (Table \ref{SI_M1}). 

Using \texttt{assess\_identifiability} and \texttt{find\_identifiable\_functions}, we can determine which parameters and which parameter combinations can be uniquely inferred from the model outputs, respectively (Table \ref{SI_M1} A). 

To perform identifiability analysis with known initial conditions, we use 
\texttt{assess\_identifiability(ode, known\_ic = [S, E, I, R])} (Table \ref{SI_M1} B).

\begin{table}[H]
\centering

\begin{minipage}[t]{0.47\linewidth}
\scriptsize
\textbf{(A)}\\[-0.3em]
\begin{verbatim}
julia> ode = @ODEmodel(
     S'(t) = - beta*S(t)*I(t)/n,
     E'(t) = (beta*S(t)*I(t)/n)- k*E(t),
     I'(t) =  k*E(t) - gamma*I(t),
     R'(t) = gamma*I(t),
     y1(t) = k*E(t)
     )

julia>  assess_identifiability(ode)
  S(t)  => :globally
  E(t)  => :globally
  I(t)  => :globally
  R(t)  => :nonidentifiable
  beta  => :nonidentifiable
  gamma => :globally
  k     => :globally
  N     => :nonidentifiable

julia> find_identifiable_functions(ode)
 k
 gamma
 N//beta
\end{verbatim}
\end{minipage}
\hspace{0.015\linewidth}
\vrule
\hspace{0.015\linewidth}
\begin{minipage}[t]{0.47\linewidth}
\scriptsize
\textbf{(B)}\\[-0.3em]
\begin{verbatim}
julia> ode = @ODEmodel(
        S'(t) = - beta*S(t)*I(t)/n,
        E'(t) = (beta*S(t)*I(t)/n)- k*E(t),
        I'(t) =  k*E(t) - gamma*I(t),
        R'(t) = gamma*I(t),
        y1(t) = k*E(t),
        y2(t) = N
        )

julia>  assess_identifiability(ode, known_ic = [S,E,I,R])
  S(0)  => :globally
  E(0)  => :globally
  I(0)  => :globally
  R(0)  => :globally
  beta  => :globally
  gamma => :globally
  k     => :globally
  N     => :globally
\end{verbatim}
\end{minipage}

\caption{Structural identifiability analysis of Model~1 (M1) using the
\texttt{StructuralIdentifiability.jl} package in Julia.
Panel~(A) shows the model input--output formulation with unknown initial conditions,
while Panel~(B) shows the corresponding formulation with known initial conditions.}
\label{SI_M1}
\end{table}

\begin{figure}[h!]
    \centering
\includegraphics[scale=1]{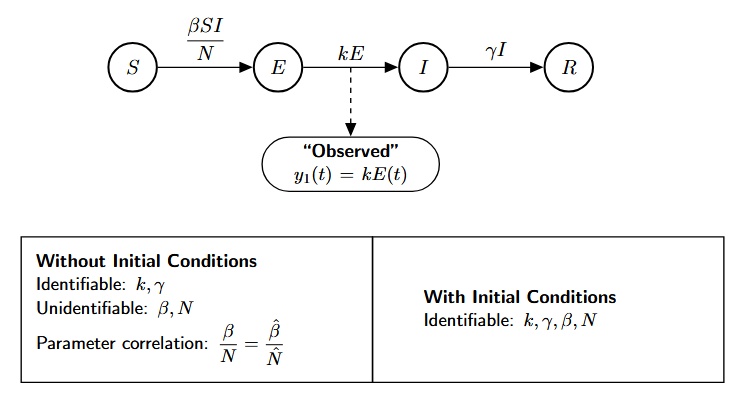}
    \caption{SEIR model flow diagram and structural identifiability results. The top panel shows the compartmental structure of the SEIR model. The bottom panel presents identifiability results from \texttt{\texttt{StructuralIdentifiability.jl}}: with unknown and with known initial conditions. Diagram reproduced with permission from Chowell et al. (2023).}
    \label{fig:M1}
\end{figure}

According to the results, the parameters $k$ and $\gamma$ are globally structurally identifiable under both scenarios. In contrast, the transmission rate $\beta$ and the total population size $N$ are not identifiable when the initial conditions are unknown. This lack of identifiability arises because $\beta$ and $N$ appear as a product in the incidence term, leading to a strong parameter correlation that hinders their separate estimation. Thus, the model is not structurally identifiable under the assumption of unknown initial conditions. However, when initial conditions are known, the total population size $N$ is also determined, allowing us to disentangle its effect from $\beta$. This enables the identifiability of both $\beta$ and $N$, rendering the model globally structurally identifiable in this case. We now formally state the result summarizing these findings.

\begin{result}
The SEIR model, as described in Model~\ref{Model1}, is not globally structurally identifiable for all parameters when only the time series of new infections, $y(t) = kE(t)$, is observed and initial conditions are unknown. In this setting, the parameters $k$ and $\gamma$ are globally identifiable, whereas $\beta$ and the total population size $N$ are not structurally identifiable due to their entanglement in the transmission term. However, when the initial conditions are known, $N$ becomes identifiable, which in turn allows for the identifiability of $\beta$, rendering the entire model globally structurally identifiable.
\end{result}

\subsection{SEIR model with symptomatic and asymptomatic
infections}

This model extends the classical SEIR framework by incorporating heterogeneity in infectiousness, distinguishing between symptomatic and asymptomatic individuals. It consists of five compartments: susceptible individuals $S(t)$, exposed (latent) individuals $E(t)$, symptomatic infectious individuals $I(t)$, asymptomatic infectious individuals $A(t)$, and recovered individuals $R(t)$. Susceptible individuals become exposed through contact with symptomatic cases at a rate of $\beta I(t)/N$. A fraction $\rho$ of exposed individuals progress to the symptomatic compartment at rate $k\rho$, while the remaining fraction $(1-\rho)$ develops asymptomatic infections at rate $k(1-\rho)$. Both infectious groups recover at the same rate $\gamma$, contributing to the recovered population.

The SEIAR model is defined by the following system of ordinary differential equations:

\begin{equation}\label{Model2}\tag{M$_2$}
	\textbf{Model 2:}
	\begin{cases}
		\displaystyle \frac{dS}{dt} = -\beta\frac{SI}{N}, \quad S(0)=S_0, \\[1.5ex]
		\displaystyle \frac{dE}{dt} = \beta\frac{SI}{N}-kE, \quad E(0)=E_0,  \\[1.5ex]
		\displaystyle \frac{dI}{dt} = k \rho E-\gamma I, \quad I(0)=I_0, \\[1.5ex]
		\displaystyle \frac{dA}{dt} = k (1-\rho) E-\gamma A, \quad A(0)=A_0, \\[1.5ex]
		\displaystyle \frac{dR}{dt} = \gamma I +  \gamma A, \quad R(0)=R_0.
	\end{cases}
\end{equation}

The structural identifiability of the model parameters is assessed using the \texttt{\texttt{StructuralIdentifiability.jl}} package. We assume the observation is the number of newly symptomatic cases, given by $y(t)=k \rho E(t)$. 

\begin{figure}[h!]
    \centering
\includegraphics[scale=0.7]{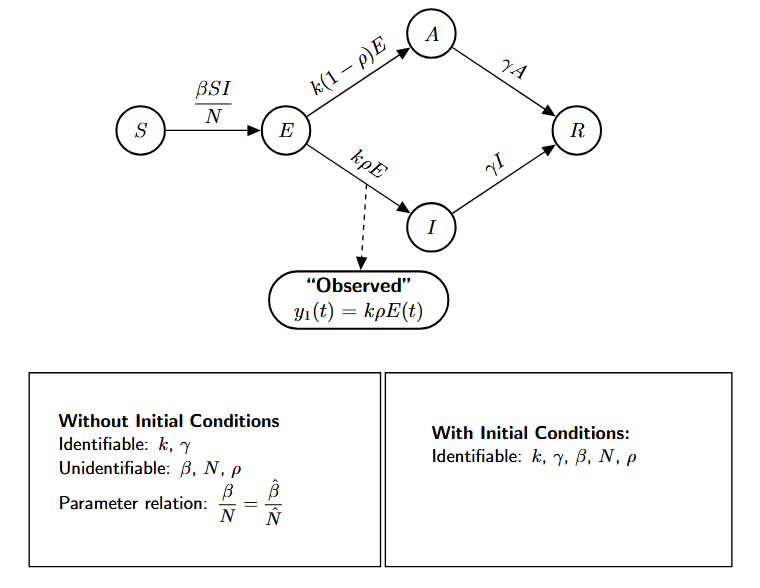}
    \caption{Flow diagram of the SEIR model extended to account for symptomatic and asymptomatic transmission dynamics. The model captures distinct progression pathways from exposed individuals to symptomatic ($I$) and asymptomatic ($A$) infectious compartments, both of which contribute to onward transmission and recovery. This enhanced structure allows for more realistic modeling of partially observed epidemics, particularly those involving subclinical spread. The bottom panel presents the structural identifiability results obtained using \texttt{StructuralIdentifiability.jl}, under both unknown and known initial condition scenarios. Diagram reproduced with permission from Chowell et al. (2023).}
    \label{fig:M2}
\end{figure}

Julia input and output for both scenarios: the unknown and known initial conditions are shown below.

\begin{table}[h!]
\centering

\begin{minipage}[t]{0.47\linewidth}
\scriptsize
\textbf{(A)}\\[-0.3em]
\begin{verbatim}
julia> ode = @ODEmodel(
    S'(t) = -beta*S(t)*I(t)/n,
    E'(t) = beta*S(t)*I(t)/n- k*E(t),
    I'(t) =  k*rho*E(t) - gamma*I(t),
    A'(t) = k*(1-rho)*E(t) - gamma*A(t),
    y1(t) = k*rho*E(t)

julia> assess_identifiability(ode)  
 S(t)  => :nonidentifiable
  E(t)  => :nonidentifiable
  I(t)  => :globally
  A(t)  => :nonidentifiable
  beta  => :nonidentifiable
  gamma => :globally
  k     => :globally
  N     => :nonidentifiable
  rho   => :nonidentifiable

julia> find_identifiable_functions(ode)
 k
 gamma
 N//beta
\end{verbatim}
\end{minipage}
\hspace{0.015\linewidth}
\vrule
\hspace{0.015\linewidth}
\begin{minipage}[t]{0.47\linewidth}
\scriptsize
\textbf{(B)}\\[-0.3em]
\begin{verbatim}
julia> ode = @ODEmodel(
    S'(t) = -beta*S(t)*I(t)/n,
    E'(t) = beta*S(t)*I(t)/n- k*E(t),
    I'(t) =  k*rho*E(t) - gamma*I(t),
    A'(t) = k*(1-rho)*E(t) - gamma*A(t),
    y1(t) = k*rho*E(t),
    y2(t) = N
        )

julia> assess_identifiability(ode, known_ic = [S,E,I,A,R])
  S(0)  => :globally
  E(0)  => :globally
  I(0)  => :globally
  A(0)  => :globally
  beta  => :globally
  gamma => :globally
  k     => :globally
  N     => :globally
  rho   => :globally
\end{verbatim}
\end{minipage}

\caption{Structural identifiability analysis of Model~2 (M2) using the
\texttt{StructuralIdentifiability.jl} package in Julia.
Panel~(A) shows the model input--output formulation with unknown initial conditions,
while Panel~(B) shows the corresponding formulation with known initial conditions.}
\label{SI_M2}
\end{table}

As in the previous case, the parameters $k$ and $\gamma$ are globally structurally identifiable regardless of the initial condition scenario. In contrast, when initial conditions are unknown, $\beta$, $N$, and $\rho$ are not identifiable. The non-identifiability of $\beta$ and $N$ stems from their coupling in the incidence term, making only their combined effect estimable. The parameter $\rho$ also remains unidentifiable under this scenario, as it cannot be uniquely recovered from the available observable $y(t) = k\rho E(t)$. Although we do not present the input-output equations explicitly in this study, previous theoretical work confirms that $\rho$ does not appear in the input-output equations \cite{Chowell2023}. When initial conditions are known, the total population size $N$ becomes a known quantity, which allows the decoupling of $\beta$ from $N$ and enables the identification of both $\beta$ and $\rho$. Thus, the full model becomes structurally identifiable under the assumption of known initial conditions.

We summarize the identifiability findings in the following result.
\begin{result}
The SEIR model with symptomatic and asymptomatic infections, described in Model~\ref{Model2}, is not structurally identifiable for all parameters when only newly symptomatic cases, $y(t) = k\rho E(t)$, are observed and initial conditions are unknown. In this setting, the parameters $\beta$, $N$, and $\rho$ are not identifiable, while $k$ and $\gamma$ remain globally structurally identifiable. The non-identifiability of $\beta$ and $N$ arises from their coupling in the transmission term, and $\rho$ cannot be uniquely determined from the observed output.

However, when initial conditions are known, the total population size $N$ becomes a known quantity, enabling the separate identification of $\beta$. Moreover, $\rho$ becomes identifiable, as its contribution to the observed output can then be disentangled. Therefore, the model becomes globally structurally identifiable under the assumption of known initial conditions.
\end{result}

\subsection{SEIR model with infectious asymptomatic individuals}

In this section, we present an extended version of the previous model by explicitly incorporating the contribution of asymptomatic individuals to the transmission process. This addition reflects growing empirical evidence that asymptomatic carriers can play a non-negligible role in the spread of infection, particularly in respiratory and emerging infectious diseases. As in Model~\ref{Model2}, exposed individuals transition to either the symptomatic or asymptomatic infectious class at rates $k\rho$ and $k(1-\rho)$, respectively. Here, $k$ denotes the rate at which individuals leave the latent period, and $\rho$ represents the fraction who become symptomatic. Importantly, both symptomatic and asymptomatic individuals are assumed to recover at the same rate $\gamma$, but now both classes contribute to onward transmission—each with its own transmission rate parameter. This enhanced structure enables a more realistic representation of heterogeneous transmission pathways in partially observed epidemics.

\begin{equation}\label{Model3}\tag{M$_3$}
\textbf{Model 3:}
\begin{cases}
\displaystyle\frac{dS}{dt} = -\frac{(\beta_A A+\beta_I I)S}{N},  \quad S(0) = S_0 \\[1.5ex]
\displaystyle \frac{dE}{dt} = \frac{(\beta_A A+\beta_I I)S}{N} - kE,  \quad E(0) = E_0\\[1.5ex]
\displaystyle \frac{dI}{dt} = k\rho E-\gamma I, \quad I(0) = I_0 \\[1.5ex]
\displaystyle \frac{dA}{dt} =k(1-\rho) E -\gamma A, \quad A(0)=A_0 \\[1.5ex]
\displaystyle \frac{dR}{dt} = \gamma I+\gamma A, \quad R(0) = R_0. 
\end{cases}
\end{equation}

The observed data remains the number of new symptomatic cases  $y = k \rho E(t).$ We perform a structural identifiability analysis similar to the previous models. The Julia input and output for both scenarios, unknown and known initial conditions, are presented below.

\begin{table}[H]
\centering

\begin{minipage}[t]{0.47\linewidth}
\scriptsize
\textbf{(A)}\\[-0.3em]
\begin{verbatim}
julia> ode = @ODEmodel(
 S'(t) =  -(betaA*A(t) + betaI*I(t))*S(t)/N,
 E'(t) = (betaA*A(t) + betaI*I(t))*S(t)/N
          - k*E(t),
 I'(t) =  k*rho1*E(t) - gamma*I(t),
 A'(t) = k*(1-rho1)*E(t) - gamma*A(t),
 R'(t) = gamma*I(t) + gamma*A(t),
 y1(t) = k*rho1*E(t)
 )

julia>  assess_identifiability(ode)
  S(t)  => :nonidentifiable
  E(t)  => :nonidentifiable
  I(t)  => :nonidentifiable
  A(t)  => :nonidentifiable
  R(t)  => :nonidentifiable
  betaA => :nonidentifiable
  betaI => :nonidentifiable
  gamma => :globally
  k     => :globally
  N     => :nonidentifiable
  rho1  => :nonidentifiable

julia> find_identifiable_functions(ode)
 k
 gamma
 (N*rho1)//(betaA*rho1 - betaA - betaI*rho1)
\end{verbatim}
\end{minipage}
\hspace{0.015\linewidth}
\vrule
\hspace{0.015\linewidth}
\begin{minipage}[t]{0.47\linewidth}
\scriptsize
\textbf{(B)}\\[-0.3em]
\begin{verbatim}
julia> ode = @ODEmodel(
    S'(t) =  -(betaA*A(t) + betaI*I(t))*S(t)/n,
    E'(t) = (betaA*A(t) + betaI*I(t))*S(t)/n- k*E(t),
    I'(t) =  k*rho1*E(t) - gamma*I(t),
    A'(t) = k*(1-rho1)*E(t) - gamma*A(t),
    R'(t) = gamma*I(t) + gamma*A(t),
    y1(t) = k*rho1*E(t),
    y2(t) = N
    )

julia> assess_identifiability(ode, known_ic = [S,E,I,A,R])
  S(0)  => :globally
  E(0)  => :globally
  I(0)  => :globally
  A(0)  => :globally
  R(0)  => :globally
  betaA => :globally
  betaI => :globally
  gamma => :globally
  k     => :globally
  N     => :globally
  rho1  => :globally


\end{verbatim}
\end{minipage}

\caption{Structural identifiability analysis of Model~3 (M3) using the
\texttt{StructuralIdentifiability.jl} package in Julia.
Panel~(A) shows the model input--output formulation with unknown initial conditions,
while Panel~(B) shows the corresponding formulation with known initial conditions.}
\label{SI_M3}
\end{table}

\begin{figure}[H]
    \centering
\includegraphics[scale=0.7]{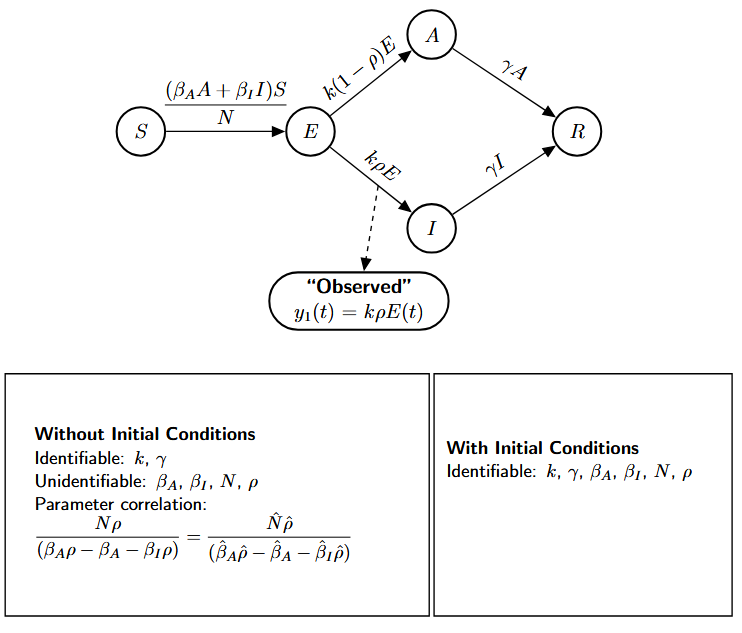}
    \caption{Flow diagram of the SEIR model with infectious asymptomatic individuals. The diagram illustrates the transitions between the susceptible, exposed, symptomatic, asymptomatic, and recovered states in the population. The bottom panel presents identifiability results from \texttt{\texttt{StructuralIdentifiability.jl}}: with unknown and with known initial conditions. Diagram reproduced with permission from Chowell et al. (2023).}
    \label{fig:M3}
\end{figure}

Based on the structural identifiability analysis, we summarize the key findings for Model~\ref{Model3} below.

\begin{result}
The SEIR model with infectious asymptomatic individuals, as described in Model~\ref{Model3}, is not structurally identifiable for all parameters when initial conditions are unknown. In this setting, the parameters $k$ and $\gamma$ are globally structurally identifiable, while $\beta_I$, $\beta_A$, $N$, and $\rho$ are not identifiable due to parameter coupling and insufficient observational information. However, when all initial conditions are known, the model becomes globally structurally identifiable, as the known initial values help disentangle the effects of individual parameters and resolve otherwise unidentifiable combinations.
\end{result}

\subsection{SEIR model with disease-induced deaths}
In the first three epidemic models, we assumed that all infected individuals eventually recover, with no disease-related mortality. In contrast, Model~\ref{Model4} extends the SEIR framework to explicitly incorporate disease-induced deaths, providing a more realistic representation of high-severity infectious diseases. This modification is captured by the following system of differential equations.

\begin{equation}\label{Model4}\tag{M$_4$}
\textbf{Model 4:}
\begin{cases}
\displaystyle\frac{dS}{dt} = -\beta\frac{SI}{N(t)},  \quad S(0)=S_0\\[1.5ex]
\displaystyle \frac{dE}{dt} = \beta\frac{SI}{N(t)} - kE, \quad E(0)=E_0  \\[1.5ex]
\displaystyle \frac{dI}{dt} = k E-(\gamma+\delta)I, \quad I(0)=I_0 \\[1.5ex]
\displaystyle \frac{dR}{dt} =\gamma I, \quad R(0)=R_0 \\[1.5ex]
\displaystyle \frac{dD}{dt} = \delta I,  \quad D(0)=D_0.
\end{cases}
\end{equation}

In this model, susceptible individuals transition to the exposed class at a rate proportional to $\beta I/N$, where $\beta$ represents the transmission rate. Exposed individuals then progress to the infectious stage at rate $k$, and infected individuals either recover at rate $\gamma$ or die due to the disease at rate $\delta$, representing disease-induced mortality. The inclusion of a mortality compartment introduces a key structural change: the total population size $N$ is no longer constant but decreases over time, reflecting the cumulative toll of the epidemic. We assess structural identifiability under two observation scenarios:  
(a) when only new symptomatic infections are observed, $y_1(t) = kE(t)$, and  
(b) when both new symptomatic infections and disease-induced deaths are observed, $y_1(t) = kE(t)$ and $y_2(t) = \delta I(t)$. \\ 

\textbf{(a) Number of new infected cases is observed:}
We perform a structural identifiability analysis following the same approach as in the previous models. The Julia input and output for the unknown and known initial condition scenarios are shown below.

\begin{table}[H]
\centering

\begin{minipage}[t]{0.47\linewidth}
\scriptsize
\textbf{(A)}\\[-0.3em]
\begin{verbatim}
julia> ode = @ODEmodel(
 S'(t) = -beta*S(t)*I(t)/(S(t)+E(t)+I(t)+R(t)),
 E'(t) = beta*S(t)*I(t)/(S(t)+E(t)+I(t)+R(t))
         -k*E(t),
 I'(t) =  k*E(t)- (gamma+delta)*I(t),
 R'(t) = gamma*I(t),
 D'(t) = delta*I(t),
 y1(t) = k*E(t)
 )

julia>  assess_identifiability(ode)
  S(t)  => :globally
  E(t)  => :globally
  I(t)  => :globally
  R(t)  => :nonidentifiable
  D(t)  => :nonidentifiable
  beta  => :nonidentifiable
  delta => :nonidentifiable
  gamma => :nonidentifiable
  k     => :globally

julia> find_identifiable_functions(ode)
 k
 delta + gamma
 delta//beta
\end{verbatim}
\end{minipage}
\hspace{0.015\linewidth}
\vrule
\hspace{0.015\linewidth}
\begin{minipage}[t]{0.47\linewidth}
\scriptsize
\textbf{(B)}\\[-0.3em]
\begin{verbatim}
julia> ode = @ODEmodel(
    S'(t) =  -beta*S(t)*I(t)/(S(t)+E(t)+I(t)+R(t)),
    E'(t) = beta*S(t)*I(t)/(S(t)+E(t)+I(t)+R(t))- k*E(t),
    I'(t) =  k*E(t)- (gamma+delta)*I(t),
    R'(t) = gamma*I(t),
    D'(t) = delta*I(t),
    y1(t) = k*E(t)
    )

julia> assess_identifiability(ode, known_ic = [S,E,I,R,D])
  S(0)  => :globally
  E(0)  => :globally
  I(0)  => :globally
  R(0)  => :globally
  D(0)  => :globally
  beta  => :globally
  delta => :globally
  gamma => :globally
  k     => :globally
\end{verbatim}
\end{minipage}

\caption{Structural identifiability analysis of Model~4a using the
\texttt{StructuralIdentifiability.jl} package in Julia.
Panel~(A) shows the model input--output formulation with unknown initial conditions,
while Panel~(B) shows the corresponding formulation with known initial conditions.}
\label{SI_M4}
\end{table}

Based on the structural identifiability analysis, we summarize the findings for Model~\ref{Model4} under the scenario where only new infections are observed:

\begin{result}
For the SEIR model with disease-induced mortality given in Eq.~\ref{Model4}, when only new infections are observed and initial conditions are unknown, only the progression rate from exposed to infectious ($k$) is globally structurally identifiable. In contrast, the remaining parameters—including the transmission rate ($\beta$), the recovery rate ($\gamma$), and the disease-induced mortality rate ($\delta$)—are not identifiable, due to insufficient observability and parameter entanglement. However, when the initial conditions are known, this additional information enables the separate identification of all model parameters, rendering the system globally structurally identifiable.
\end{result}

\begin{figure}[H]
    \centering
\includegraphics[scale=0.7]{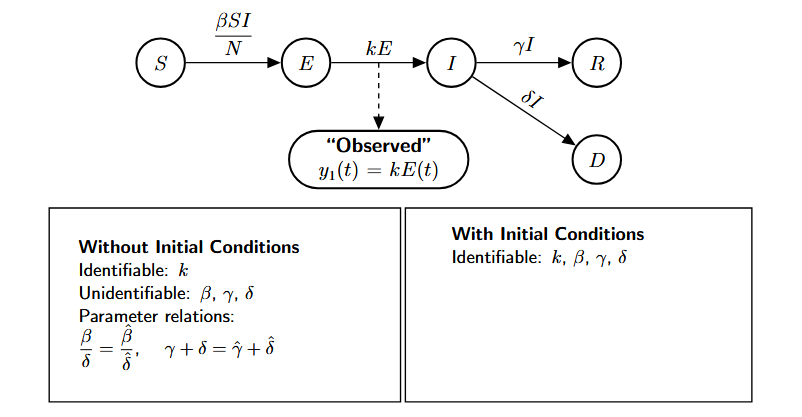}
    \caption{Flow diagram of the SEIR model extended to include disease-induced mortality. In this formulation, infected individuals may either recover or die as a direct consequence of infection, leading to a decline in total population size over time. This feature enhances the model's realism for high-severity pathogens. The bottom panel displays the structural identifiability results obtained using \texttt{StructuralIdentifiability.jl} under scenarios with both unknown and known initial conditions. Diagram reproduced with permission from Chowell et al. (2023).}
    \label{fig:M4}
\end{figure}

We now extend the analysis to the case where both the number of new infections and the number of disease-induced deaths are observed. \\

\textbf{(b) Number of new infected cases and deaths are observed:}

\begin{table}[H]
\centering

\begin{minipage}[t]{0.47\linewidth}
\scriptsize
\textbf{(A)}\\[-0.3em]
\begin{verbatim}
julia> ode = @ODEmodel(
    S'(t) =  -beta*S(t)*I(t)/(S(t)+E(t)+I(t)
             +R(t)),
    E'(t) = beta*S(t)*I(t)/(S(t)+E(t)+I(t)
            +R(t))-k*E(t),
    I'(t) =  k*E(t)- (gamma+delta)*I(t),
    R'(t) = gamma*I(t),
    D'(t) = delta*I(t),
    y1(t) = k*E(t),
    y2(t) = delta*I(t)
    )

julia>  assess_identifiability(ode)
  S(t)  => :globally
  E(t)  => :globally
  I(t)  => :globally
  R(t)  => :globally
  D(t)  => :nonidentifiable
  beta  => :globally
  delta => :globally
  gamma => :globally
  k     => :globally


julia> find_identifiable_functions(ode)
 k
 gamma
 delta
 beta
\end{verbatim}
\end{minipage}
\hspace{0.015\linewidth}
\vrule
\hspace{0.015\linewidth}
\begin{minipage}[t]{0.47\linewidth}
\scriptsize
\textbf{(B)}\\[-0.3em]
\begin{verbatim}
julia> ode = @ODEmodel(
    S'(t) =  -beta*S(t)*I(t)/(S(t)+E(t)+I(t)+R(t)),
    E'(t) = beta*S(t)*I(t)/(S(t)+E(t)+I(t)+R(t))- k*E(t),
    I'(t) =  k*E(t)- (gamma+delta)*I(t),
    R'(t) = gamma*I(t),
    D'(t) = delta*I(t),
    y1(t) = k*E(t),
    y2(t) = delta*I(t)
    )

julia> assess_identifiability(ode, known_ic = [S,E,I,R,D])

  S(0)  => :globally
  E(0)  => :globally
  I(0)  => :globally
  R(0)  => :globally
  D(0)  => :globally
  beta  => :globally
  delta => :globally
  gamma => :globally
  k     => :globally
\end{verbatim}
\end{minipage}

\caption{Structural identifiability analysis of Model~4b using the
\texttt{StructuralIdentifiability.jl} package in Julia.
Panel~(A) shows the model input--output formulation with unknown initial conditions,
while Panel~(B) shows the corresponding formulation with known initial conditions.}
\label{SI_M4b}
\end{table}

According to the results, all parameters in Model~\ref{Model4} are globally structurally identifiable when both the number of new infections and disease-induced deaths are observed, regardless of whether the initial conditions are known or unknown. These findings underscore a key methodological insight: incorporating additional observational data can substantially enhance a model’s identifiability, even in settings with partial knowledge of initial conditions. We summarize these findings in the following result.

\begin{result}
The SEIR model with disease-induced mortality, as defined in Model~\ref{Model4}, is globally structurally identifiable when both the number of new infections and the number of disease-induced deaths are observed. This identifiability holds regardless of whether the initial conditions are known or unknown. Notably, the inclusion of the second observable $y_2(t)$ allows for full parameter identifiability even in the absence of known initial conditions, highlighting a key methodological insight: augmenting the system with complementary observational data can resolve parameter confounding that would otherwise persist. These results emphasize the critical role of incorporating multiple, complementary observational data streams to ensure the reliable estimation of key epidemiological parameters.
\end{result}

\begin{figure}[H]
    \centering
\includegraphics[scale=0.7]{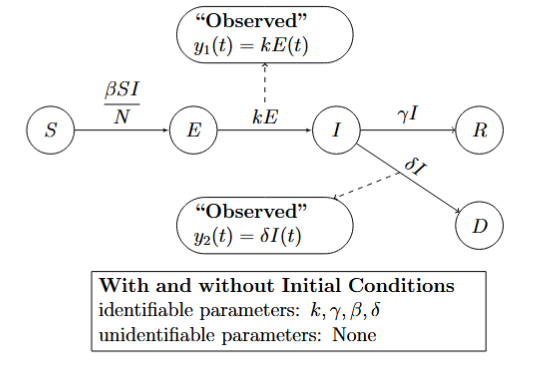}
    \caption{Flow diagram of the SEIR model incorporating disease-induced mortality and dual observational outputs. In this formulation, infected individuals may recover or die from the disease, resulting in a declining population size over time. This structure allows for the analysis of epidemics where mortality is a significant component of the disease burden. The bottom panel shows structural identifiability results obtained using \texttt{StructuralIdentifiability.jl}, considering two types of observations: newly symptomatic infections and disease-induced deaths, under both unknown and known initial condition scenarios. Diagram reproduced with permission from Chowell et al. (2023).}
    \label{fig:M4b}
\end{figure}

\subsection{Simple vector-borne disease model}

Next, we explore a class of epidemic models specifically designed to capture the transmission dynamics of vector-borne diseases, where pathogens are transmitted between humans and vector populations such as mosquitoes or ticks. This model includes five epidemiological compartments: susceptible mosquitoes $Sv(t)$, infected mosquitoes $Iv(t)$, susceptible humans $S(t)$, infected humans $I(t)$, and recovered humans $R(t)$. The mosquito population is modeled with constant recruitment at rate $\Lambda_v$ and per capita mortality at rate $\mu_v$. \\ 

Susceptible mosquitoes become infected through contact with infectious humans at a rate proportional to $\beta I(t)/N$, while susceptible humans acquire infection from infectious mosquitoes at a rate of $\beta_v I_v(t)/N$. As in previous models, infected humans recover at rate $\gamma$. Importantly, the total human population $N$ is assumed to remain constant over the course of the epidemic, while the mosquito population is dynamically regulated through birth and death processes.

\begin{equation}\label{Model5}\tag{M$_5$}
	\textbf{Model 5:}
	\begin{cases}
		\displaystyle \frac{dS_{\upsilon}}{dt} = \Lambda_{\upsilon} - \frac{\beta S_{\upsilon}I}{N} -\mu_{\upsilon} S_\upsilon, \quad S_\upsilon(0) = S_{\upsilon0}\\[1.5ex]
		\displaystyle \frac{dI_{\upsilon}}{dt} 
= \frac{\beta S_{\upsilon}I}{N}-\mu_{\upsilon} I_{\upsilon}, \quad I_{\upsilon}(0) =I_{\upsilon0}  \\[1.5ex]
		\displaystyle \frac{dS}{dt} 
=  -\frac{\beta_\upsilon S I_{\upsilon}}{N}, \quad S(0) = S_0 \\[1.5ex]
		\displaystyle \frac{dI}{dt} 
=  \frac{\beta_\upsilon S I_{\upsilon}}{N} - \gamma I, \quad I(0)=I_0 \\[1.5ex]
		\displaystyle \frac{dR}{dt} 
=  \gamma I, \quad R(0) = R_0.
	\end{cases}
\end{equation}

In this setting, the model output corresponds to the cumulative number of new human infections, defined as  
$y_1(t) = \int_{0}^{t} \frac{\beta_\upsilon S(t) I_{\upsilon}(t)}{N} \, dt = S(0) - S(t)$.  
This formulation reflects the total number of infections generated by contact with infectious mosquitoes over time. Note that in implementation of the identifiability analysis, we let $S(0)$ be represented by the constant $c\in\mathbb{R}$, which is treated as an unknown parameter.\\ 

We now assess whether Model~\ref{Model5} is structurally capable of revealing its underlying epidemiological parameters from this cumulative incidence measure, under the assumptions of the model structure and observational setup.

\begin{table}[H]
\centering

\begin{minipage}[t]{0.47\linewidth}
\scriptsize
\textbf{(A)}\\[-0.3em]
\begin{verbatim}
julia> ode = @ODEmodel(
 Sv'(t) = lamdaupsilon - beta*Sv(t)*I(t)/n- 
          muupsilon*Sv(t),
 Iv'(t) = beta*Sv(t)*I(t)/N - muupsilon*Iv(t),
 S'(t) = -betaupsilon*Iv(t)*S(t)/N,
 I'(t) = betaupsilon*Iv(t)*S(t)/N - gamma*I(t),
 R'(t) = gamma*I(t),
 y1(t) = c-S(t)
            )
                     
  Sv(t)        => :nonidentifiable
  Iv(t)        => :nonidentifiable
  S(t)         => :globally
  I(t)         => :globally
  R(t)         => :nonidentifiable
  beta         => :nonidentifiable
  betaupsilon  => :nonidentifiable
  c            => :globally
  gamma        => :globally
  lamdaupsilon => :nonidentifiable
  muupsilon    => :globally
  N            => :nonidentifiable

julia> find_identifiable_functions(ode)

 muupsilon
 gamma
 c
 N//beta
 N//(betaupsilon*lamdaupsilon)

\end{verbatim}
\end{minipage}
\hspace{0.015\linewidth}
\vrule
\hspace{0.015\linewidth}
\begin{minipage}[t]{0.47\linewidth}
\scriptsize
\textbf{(B)}\\[-0.3em]
\begin{verbatim}
julia> ode = @ODEmodel(
 Sv'(t) = lamdaupsilon - beta*Sv(t)*I(t)/n- 
          muupsilon*Sv(t),
 Iv'(t) = beta*Sv(t)*I(t)/N - muupsilon*Iv(t),
 S'(t) = -betaupsilon*Iv(t)*S(t)/N,
 I'(t) = betaupsilon*Iv(t)*S(t)/N - gamma*I(t),
 R'(t) = gamma*I(t),
 y1(t) = c-S(t),
 y2(t) = N,
 y3(t) = c
         )

julia> assess_identifiability(ode, known_ic = [Sv,Iv,S,I,R])

  Sv(0)        => :globally
  Iv(0)        => :globally
  S(0)         => :globally
  I(0)         => :globally
  R(0)         => :globally
  beta         => :globally
  betaupsilon  => :globally
  gamma        => :globally
  lamdaupsilon => :globally
  muupsilon    => :globally
  N            => :globally
\end{verbatim}
\end{minipage}

\caption{Structural identifiability analysis of Model~5 using the
\texttt{StructuralIdentifiability.jl} package in Julia. Here, the constant $c$ represents the initial susceptible population $S(0)$, treated as an unknown parameter in the symbolic model. Thus, the observable 
$y_1(t)=S(0)-S(t)$ is implemented as 
$y_1(t)=c-S(t)$.
Panel~(A) shows the model input--output formulation with unknown initial conditions,
while Panel~(B) shows the corresponding formulation with known initial conditions.}
\label{SI_M5}
\end{table}

\begin{figure}[H]
    \centering
\includegraphics[scale=0.7]{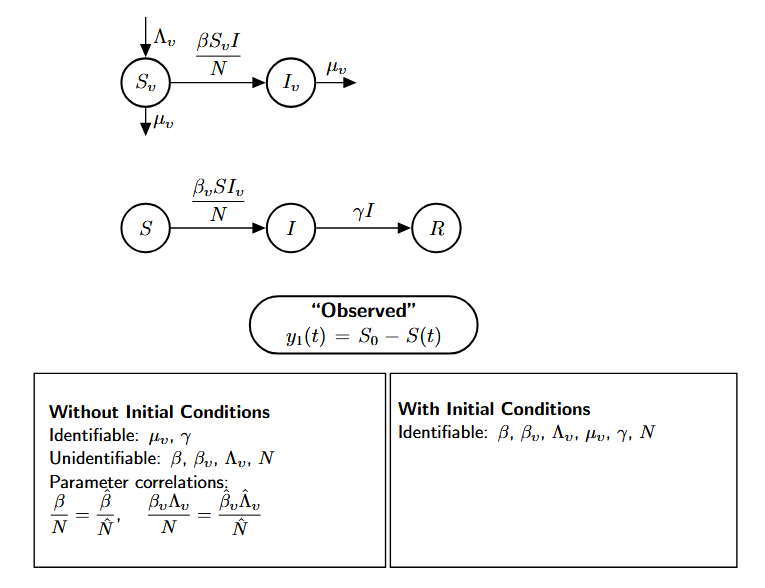}
    \caption{Flow diagram of the simple vector-borne disease model. This model captures the transmission dynamics between human and mosquito populations. Susceptible mosquitoes become infected through contact with infectious humans, while susceptible humans acquire infection from infectious mosquitoes. Infected humans eventually recover, while mosquitoes do not recover but are removed through natural mortality. The bottom panel presents structural identifiability results obtained using \texttt{StructuralIdentifiability.jl}, under both unknown and known initial condition scenarios, highlighting how prior knowledge about initial states influences parameter identifiability. Diagram reproduced with permission from Chowell et al. (2023).}
    \label{fig:M5}
\end{figure}

We summarize the results of the identifiability analysis below:

\begin{result}
The vector-borne disease model described in Model~\ref{Model5} is not structurally identifiable when initial conditions are unknown. In this case, only the mosquito mortality rate ($\mu_v$) and the human recovery rate ($\gamma$) are globally structurally identifiable, while the parameters $\beta$, $\beta_v$, $\Lambda_v$, and $N$ are not identifiable due to parameter coupling and limited observability. However, when all initial conditions are known, all parameters become structurally identifiable, indicating that prior knowledge of initial states plays a critical role in resolving parameter identifiability in vector-host transmission models.
\end{result}

\subsection{Vector-borne disease model with asymptomatic infections}

We now perform a structural identifiability analysis of the extended vector-borne disease model \ref{Model6}, which incorporates an asymptomatic class in the human host population—building upon the structure of Model~\ref{Model5}. This model includes six epidemiological compartments: susceptible mosquitoes $Sv(t)$, infected mosquitoes $Iv(t)$, susceptible humans $S(t)$, symptomatic infected humans $I(t)$, asymptomatic infected humans $A(t)$, and recovered humans $R(t)$.

Susceptible mosquitoes become infected after biting either symptomatic or asymptomatic humans, at rates proportional to $\beta_I$ and $\beta_A$, respectively. Infected mosquitoes die at a per capita rate $\mu_v$. On the human side, susceptible individuals become infected by infectious mosquitoes at a rate $\beta_v$. Upon infection, a proportion $\rho$ of individuals progress to the symptomatic class, while the remaining $1-\rho$ enter the asymptomatic class. Both groups recover at the same rate $\gamma$.

The total human population is assumed constant, while the mosquito population is governed by constant recruitment $\Lambda_v$ and mortality $\mu_v$, allowing for population turnover in the vector compartment.

\begin{equation}\label{Model6}\tag{M$_6$}
	\textbf{Model 6:}
	\begin{cases}
	\displaystyle \frac{dS_{\upsilon}}{dt} 
= \Lambda_{\upsilon} - \frac{\beta_{I} S_{\upsilon}I+\beta_{A} S_{\upsilon} A}{N} -\mu_{\upsilon}S_{\upsilon}, \quad S_\upsilon(0) = S_{\upsilon0}\\[1.5ex]
	\displaystyle \frac{dI_{\upsilon}}{dt} 
= \frac{\beta_{I} S_{\upsilon}I+\beta_{A} S_{\upsilon} A}{N}-\mu_{\upsilon} I_{\upsilon}, \quad I_\upsilon(0) = I_{\upsilon0}\\[1.5ex]
	\displaystyle \frac{dS}{dt} 
= - \frac{\beta_{\upsilon} I_{\upsilon}S}{N}, \quad S(0) = S_0\\[1.5ex]
	\displaystyle \frac{dI}{dt} 
= \frac{ \rho \beta_{\upsilon} I_{\upsilon}S}{N}-\gamma I, \quad I(0) = I_0\\[1.5ex]
	\displaystyle \frac{dA}{dt} 
=   \frac{(1-\rho)\beta_{\upsilon}I_{\upsilon}S }{N}  -\gamma A, \quad A(0) = A_0\\[1.5ex]
	\displaystyle \frac{dR}{dt} 
=  \gamma I +\gamma A, \quad R(0) = R_0.\\[1.5ex]

	\end{cases}
\end{equation}

The cumulative number of new symptomatic infections, denoted by $$y_1(t) =\int_{0}^{t} \frac{\rho \beta_\upsilon S I_{\upsilon}}{N}=\rho(S(0)-S(t)),$$ is taken as the observable output for the identifiability analysis. Next, we investigate whether Model \ref{Model6} can reveal its epidemiological parameters from these cumulative case data. When conducting the identifiability analysis, we let the initial condition $S(0)$ be treated as an unknown parameter that is represented by a constant parameter $c\in\mathbb{R}$ as in \ref{Model5}.

\begin{figure}[H]
    \centering
\includegraphics[scale=0.7]{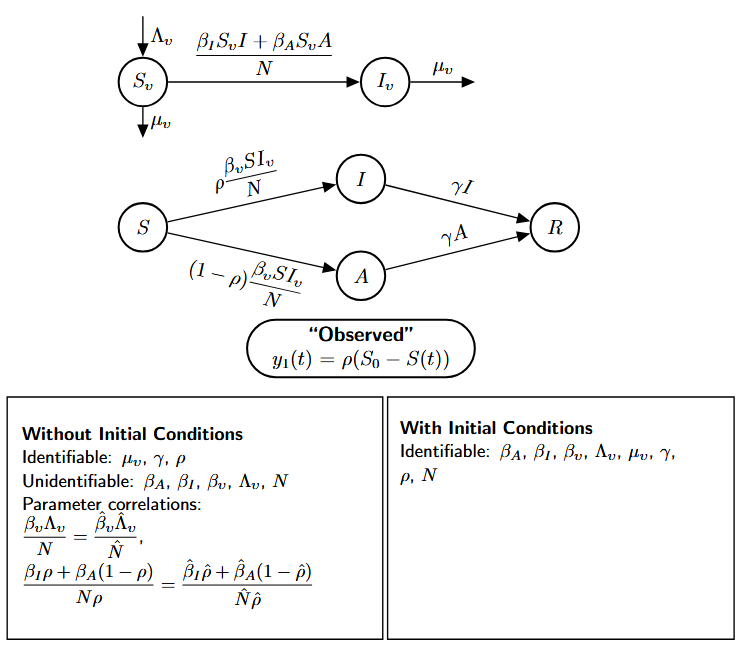}
    \caption{Flow diagram of the vector-borne disease model incorporating asymptomatic human infections. The model captures transmission pathways from both symptomatic and asymptomatic individuals to mosquitoes, and from infected mosquitoes to susceptible humans. This framework reflects more realistic vector-host dynamics in which asymptomatic individuals contribute to ongoing transmission. The bottom panel presents structural identifiability results obtained using \texttt{StructuralIdentifiability.jl}, under both unknown and known initial condition scenarios, highlighting the influence of observational assumptions on parameter identifiability. Diagram reproduced with permission from Chowell et al. (2023).}
    \label{fig:M6}
\end{figure}

\begin{table}[H]
\centering

\begin{minipage}[t]{0.47\linewidth}
\scriptsize
\textbf{(A)}\\[-0.3em]
\begin{verbatim}
julia> ode = @ODEmodel(
 Sv'(t) = lambdaupsilon-((betaI*Sv(t)*I(t)
          +betaA*Sv(t)*A(t))/N)-muv*Sv(t),
 Iv'(t) = (betaI*Sv(t)*I(t)+betaA*Sv(t)*
          A(t))/N-muv*Iv(t),
 S'(t) =  -betav*Iv(t)*S(t)/N,
 I'(t) = rho*betav*Iv(t)/N*S(t)-gamma*I(t),
 A'(t) = (1-rho)*betav*Iv(t)*S(t)/N- gamma*A(t),
 R'(t) = gamma*I(t)+ gamma*A(t),
 y1(t) =  rho*c-rho*S(t)
 )

julia>  assess_identifiability(ode)
  Sv(t)         => :nonidentifiable
  Iv(t)         => :nonidentifiable
  S(t)          => :nonidentifiable
  I(t)          => :nonidentifiable
  A(t)          => :nonidentifiable
  R(t)          => :nonidentifiable
  N             => :nonidentifiable
  betaA         => :nonidentifiable
  betaI         => :nonidentifiable
  betav         => :nonidentifiable
  c             => :nonidentifiable
  gamma         => :globally
  lambdaupsilon => :nonidentifiable
  muv           => :globally
  rho           => :nonidentifiable

julia> find_identifiable_functions(ode)
 muv
 gamma
 c*rho
 (betav*lambdaupsilon)//N
 (N*rho)//(betaA*rho - betaA - betaI*rho)
\end{verbatim}
\end{minipage}
\hspace{0.015\linewidth}
\vrule
\hspace{0.015\linewidth}
\begin{minipage}[t]{0.47\linewidth}
\scriptsize
\textbf{(B)}\\[-0.3em]
\begin{verbatim}



julia> ode = @ODEmodel(
 Sv'(t) = lambdaupsilon-((betaI*Sv(t)*I(t)
          +betaA*Sv(t)*A(t))/N)-muv*Sv(t),
 Iv'(t) = (betaI*Sv(t)*I(t)+betaA*Sv(t)*A(t))/N
          -muv*Iv(t),
 S'(t) =  -betav*Iv(t)*S(t)/N,
 I'(t) = rho*betav*Iv(t)/N*S(t)-gamma*I(t),
 A'(t) = (1-rho)*betav*Iv(t)*S(t)/N- gamma*A(t),
 R'(t) = gamma*I(t)+ gamma*A(t),
 y1(t) =  rho*5-rho*S(t),
 y2(t) = N,
 y3(t) = c
 )

julia> assess_identifiability(ode, known_ic = [Sv,Iv,S,I,A,R])
  Sv(0)         => :globally
  Iv(0)         => :globally
  S(0)          => :globally
  I(0)          => :globally
  A(0)          => :globally
  R(0)          => :globally
  betaA         => :globally
  betaI         => :globally
  betav         => :globally
  gamma         => :globally
  lambdaupsilon => :globally
  muv           => :globally
  N             => :globally
  rho           => :globally
\end{verbatim}
\end{minipage}

\caption{Structural identifiability analysis of Model~6 using the
\texttt{StructuralIdentifiability.jl} package in Julia. Here, the constant $c$ represents the initial susceptible population $S(0)$, treated as an unknown parameter in the symbolic model. Thus, the observable 
$y_1(t)=\rho (S(0)-S(t))$ is implemented as 
$y_1(t)=\rho c-\rho S(t)$.
Panel~(A) shows the model input--output formulation with unknown initial conditions,
while Panel~(B) shows the corresponding formulation with known initial conditions.}
\label{SI_M6}
\end{table}

Based on the structural identifiability findings summarized in Figure~\ref{fig:M6} and Table~\ref{SI_M6}, we formally state the following result.

\begin{result}
The vector-borne disease model \ref{Model6}, which incorporates asymptomatic infections in humans, is not structurally identifiable when initial conditions are unknown. Under this scenario, only the parameters $\mu_\upsilon$ (mosquito mortality rate) and $\gamma$ (human recovery rate) are globally structurally identifiable. In contrast, key transmission parameters—$\beta_I$, $\beta_A$, $\beta_\upsilon$, $\Lambda_\upsilon$, $N$, and $\rho$ (proportion of symptomatic cases)—remain unidentifiable. However, when all initial conditions are known, the model becomes fully structurally identifiable, underscoring the importance of initial state information for parameter resolution in vector-host systems.
\end{result}

\subsection{Ebola model}

This epidemic model characterizes the transmission dynamics of Ebola virus disease (EVD), a pathogen with multiple transmission pathways and clinical outcomes. The model includes six epidemiological compartments: susceptible individuals $S(t)$, latent individuals $E(t)$, infected individuals in the community $I(t)$, hospitalized individuals $H(t)$, recovered individuals $R(t)$, and disease-induced deaths $D(t)$. \\

Susceptible individuals can become exposed by contact with infectious individuals in the community at a rate $\beta_I I(t)/N$, hospitalized individuals at $\beta_H H(t)/N$, or deceased individuals at $\beta_D D(t)/N$. Following exposure, individuals transition to the infectious class after a mean latent period of $1/k$ days. Infectious individuals in the community can recover at rate $\gamma_I$ or die at rate $\delta_I$, while hospitalized individuals recover at rate $\gamma_H$ or die at rate $\delta_H$. \\ 

We aim to evaluate the structural identifiability of Model~\ref{Model7} under three distinct observation scenarios:  
a) only new infection cases are observed,  
b) new infections and hospitalizations are observed,  
c) new infections, hospitalizations, and deaths are jointly observed. \\ 

The complete system of differential equations representing this model is presented below.

\begin{equation}\label{Model7}\tag{M$_7$}
\textbf{Model 7:}
\begin{cases}
\displaystyle\frac{dS}{dt}= \displaystyle\frac{-(\beta_{I} I + \beta_{H} H+\beta_{D} D) S}{N}, \quad S(0) = S_0\\[1.5ex]
 \displaystyle\frac{dE}{dt}=  \displaystyle\frac{(\beta_{I} I + \beta_{H} H+\beta_{D} D) S}{N}-kE, \quad E(0) = E_0\\[1.5ex]
\displaystyle\frac{dI}{dt}=kE-(\alpha+\gamma_{I}+\delta_{I})I, \quad I(0) = I_0\\[1.5ex]
\displaystyle\frac{dH}{dt}=  \alpha I -(\gamma_{H}+\delta_{H}) H,\quad H(0) = H_0\\[1.5ex]  
\displaystyle\frac{dR}{dt}=  \gamma_I I +\gamma_{H} H, \quad R(0) = R_0\\[1.5ex]
 \displaystyle\frac{dD}{dt}=  \delta_{I} I+\delta_{H} H, \quad D(0) = D_0.\\[1.5ex]
 \end{cases}
\end{equation}

\begin{figure}[H]
    \centering
\includegraphics[scale=0.9]{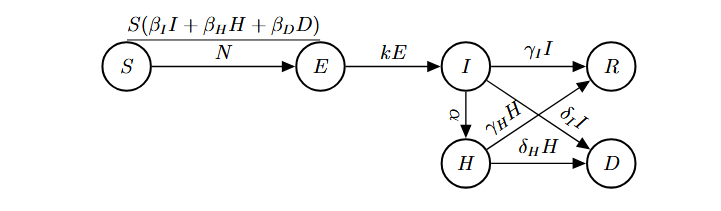}
    \caption{Flow diagram of the Ebola transmission model. The model captures multiple transmission pathways, including exposure to the virus via contact with infectious individuals in the community, hospitalized patients, and deceased individuals, each with distinct transmission rates. This structure reflects the complex and high-risk nature of Ebola virus transmission, particularly in the presence of unsafe burial practices and limited healthcare infrastructure. Diagram reproduced with permission from Chowell et al. (2023).}
    \label{fig:M7}
\end{figure}

We begin our identifiability assessment by considering the scenario in which only new infection cases are observed, corresponding to a minimal observational setting. \\ 

\textbf{(a) Number of new infected cases is observed}:Due to the high dimensionality and algebraic complexity of the model, the structural identifiability analysis using only new infection cases as observations could not be completed with the \texttt{StructuralIdentifiability.jl} package in Julia. Similar computational challenges were observed with other symbolic tools, including DAISY \cite{Chowell2023}. To overcome these limitations, we employed a model reduction strategy based on first integrals prior to identifiability analysis, although in the examples considered here these integrals were derived analytically on a model-specific basis rather than through a fully automated procedure.

\subsubsection{Reducing the model using first integral technique:}

When structural identifiability analysis becomes computationally infeasible due to model complexity, a practical solution is to simplify the system using generalized first integrals. This reduction technique preserves the model’s essential dynamical structure while improving the tractability for identifiability analysis.

Symbolic approaches based on input-output equations—such as those implemented in the \texttt{StructuralIdentifiability.jl} package—typically involve two key steps:
\begin{enumerate}
    \item Differential elimination, that is, generating input-output equations that relate observables and parameters.
    \item Identifiability assessment, by analyzing the coefficients of these equations to determine identifiable parameters or combinations.
\end{enumerate}
In many cases, differential elimination becomes the main computational bottleneck. To alleviate this challenge, a useful strategy is to trade states for parameters—reducing the number of dynamic variables in exchange for introducing constants of integration. While this may complicate the second step (coefficient analysis), it often enables otherwise intractable models to be analyzed. The following examples illustrate this technique.

\subsubsection*{Example 1: Model with a linear first integral}

Consider the following system:
\begin{equation}\label{ex1-original}
\begin{cases}
x_1' = \mu x_1^2 + x_2^2, \\
x_2' = -\mu x_1^2 - x_2^2, \\
y = x_1.
\end{cases}
\end{equation}

Observe that the sum of the derivatives satisfies
\[
(x_1 + x_2)' = 0.
\]
Integrating both sides of the equation above yields $x_1 + x_2 = C$ for some constant $C$. We will refer to this as the first integral. We can use this first integral to eliminate $x_2$ by plugging it into the first equation in \ref{ex1-original}, yielding the reduced system:
\begin{equation}\label{ex1-reduced}
\begin{cases}
x_1' = \mu x_1^2 + (C - x_1)^2, \\
y = x_1.
\end{cases}
\end{equation}

The input-output equation for the reduced model \eqref{ex1-reduced} can be readily derived by substituting $x_1 = y$. Because the transformation from the original system \eqref{ex1-original} to the reduced system \eqref{ex1-reduced} is invertible and one-to-one, the identifiability properties of shared elements—such as the parameter $\mu$ and state $x_1$—are fully preserved. Furthermore, if both $x_1$ and the constant $C$ can be shown to be identifiable in the reduced system, then $x_2 = C - x_1$ must also be identifiable in the original model. This confirms that identifiability can be reliably inferred from the reduced system under a valid transformation.

\vspace{1em}
\subsubsection*{Example 2: Model with a generalized first integral}

Now consider a slightly modified system:
\begin{equation}\label{ex2-original}
\begin{cases}
x_1' = \mu_1 x_2^2, \\
x_2' = -\mu_2 x_1^2 - x_2^2, \\
y = x_1.
\end{cases}
\end{equation}

This model does not admit an obvious first integral, but we can derive the following relation:
\[
x_1' + \mu_1 x_2' + \mu_1 \mu_2 y^2 = 0,
\]

which implies that there exists a constant $C_1$ such that:
\[
x_1 + \mu_1 x_2 + \mu_1 \mu_2 \int_0^t y^2(\tau)\, d\tau = C_1.
\]

To use this for reduction, we introduce an auxiliary state $x_3$ defined by $x_3' = x_1^2$ and an additional output $y_2 = x_3$, which produces:
\begin{equation}\label{ex2-extended}
\begin{cases}
x_1' = \mu_1 x_2^2, \\
x_2' = -\mu_2 x_1^2 - x_2^2, \\
x_3' = x_1^2, \\
y = x_1, \quad y_2 = x_3.
\end{cases}
\end{equation}

This formulation introduces $x_3 = C_2 + \int_0^t y^2(\tau)\, d\tau$, and since $x_3$ is observed, the constant $C_2$ becomes identifiable. The system now admits a first integral:
\[
x_1 + \mu_1 x_2 + \mu_1 \mu_2 x_3 = C,
\]
which allows us to eliminate $x_2$ as $x_2 = (C - x_1 - \mu_1 \mu_2 x_3)/\mu_1$. Substituting this into the system, we get:
\begin{equation}\label{ex2-reduced}
\begin{cases}
x_1' = \mu_1 \left( \frac{C - x_1 - \mu_1 \mu_2 x_3}{\mu_1} \right)^2, \\
x_3' = x_1^2, \\
y = x_1, \quad y_2 = x_3.
\end{cases}
\end{equation}

The reduced system \eqref{ex2-reduced} retains the same number of dynamic states as the original model \eqref{ex2-original}, but the inclusion of two output variables significantly facilitates the derivation of input-output equations. As in the previous example, the identifiability properties are preserved because no new parameters were introduced, and the transformation remains invertible and information-preserving.

\vspace{1em}

This model reduction strategy based on generalized first integrals was originally introduced in \cite{Pogudin2024}, where it demonstrated substantial computational gains in the structural identifiability analysis of chemical reaction networks. It has since been recommended by \cite{Ballif2024} for enabling the identifiability analysis of more complex biological systems that would otherwise be analytically intractable. In the present work, first integrals are derived analytically and used to reduce models in cases where direct structural identifiability analysis of the original system does not yield conclusive results.

Building on these insights, we apply the generalized first-integral reduction technique to simplify Model~\ref{Model7}, allowing us to conduct structural identifiability analysis in the presence of the computational limitations encountered with the full, unreduced system.

The original model, including the observation equation, is specified below:

\begin{align*}
    S'(t) &= -\frac{(\beta_I I + \beta_H H + \beta_D D) S}{S + E + I + H + R}, \\
    E'(t) &= \frac{(\beta_I I + \beta_H H + \beta_D D)S}{S + E + I + H + R} - k E, \\
    I'(t) &= k E - (\alpha + \gamma_I + \delta_I) I, \\
    H'(t) &= \alpha I - (\gamma_H + \delta_H) H, \\
    R'(t) &= \gamma_I I + \gamma_H H, \\
    D'(t) &= \delta_I I + \delta_H H, \\
    y(t) &= k E.
\end{align*}
Since $R(t)$ does not appear in any equation except as part of the total population $N(t) = S(t) + E(t) + I(t) + H(t) + R(t)$, we eliminate $R(t)$. We observe that,
\[
S'(t) + E'(t) + I'(t) + H'(t) + R'(t) + D'(t) = 0 \Rightarrow S(t) + E(t) + I(t) + H(t) + R(t) + D(t) = C,
\]
We define $N(t) = C - D(t)$. Also, from the model, we obtain
\[
S'(t) + E'(t) + y(t) = 0 \Rightarrow S(t) + E(t) + \int_0^t y(\tau)\, d\tau = C_2.
\]
Let us define a new variable:
\[
x_1(t) = \int_0^t y(\tau)\, d\tau = \int_0^t k E(\tau)\, d\tau,
\]
and hence,
\[
S(t) = C_2 - E(t) - x_1(t).
\]
Next, we added the integral (of the output) to the model as a new state variable and new observation (since the integral can be observed), and we use $S(t) = C_2 - E(t) - x_1(t)$ to eliminate $S(t)$, the model with observations becomes,

\begin{align*}
    E'(t) &= \frac{(\beta_I I(t) + \beta_H H(t) + \beta_D D(t))(C_2 - E(t) - \int_0^t y(\tau))}{C - D(t)} - k E(t), \\
    I'(t) &= k E(t) - (\alpha + \gamma_I + \delta_I) I(t), \\
    H'(t) &= \alpha I(t) - (\gamma_H + \delta_H) H(t), \\
    D'(t) &= \delta_I I(t) + \delta_H H(t), \\
    x_1'(t) &= k E(t), \\
    y(t) &= k E(t), \\
    y_2(t) &= x_1(t).
\end{align*}
Next, from the original equations for $H'(t)$ and $D'(t)$, solve for $I(t)$:

\[
I(t) = \frac{\delta_H H'(t) + (\gamma_H + \delta_H) D'(t)}{\alpha \delta_H + (\gamma_H + \delta_H)} \cdot \frac{1}{\delta_I}.
\]
From the equation for $I'(t)$,
\[
I'(t) = y(t) - (\alpha + \gamma_I + \delta_I) I(t),
\]
substitute the expression for $I(t)$,
\[
I'(t) = y(t) - (\alpha + \gamma_I + \delta_I) \cdot \left( \frac{\delta_H H'(t) + (\gamma_H + \delta_H) D'(t)}{\alpha \delta_H + (\gamma_H + \delta_H)} \cdot \frac{1}{\delta_I} \right).
\]
Integrating both sides,
\[
I(t) = \int_0^t y(\tau) - (\alpha + \gamma_I + \delta_I) \cdot \left( \frac{\delta_H H(t) + (\gamma_H + \delta_H) D(t)}{\alpha \delta_H + (\gamma_H + \delta_H)} \cdot \frac{1}{\delta_I} \right) + C_3,
\]
which implies,
\[
\int_0^t y(\tau) = I(t) + (\alpha + \gamma_I + \delta_I) \cdot \left( \frac{\delta_H H(t) + (\gamma_H + \delta_H) D(t)}{\alpha \delta_H + (\gamma_H + \delta_H)} \cdot \frac{1}{\delta_I} \right) - C_3.
\]
Substitute $\int_0^t y(\tau)$ into the model again and obtain
\begin{align*}
 E'(t) &= \dfrac{(\beta_I I(t) + \beta_H H(t) + \beta_D D(t)) \left(C_2 - E(t) - I(t) - (\alpha + \gamma_I + \delta_I)  \cdot \epsilon(t)+ C_3 \right)}{C - D(t)} -  k E(t), \\
 & \text{where } \epsilon(t)=\left( \frac{\delta_H H(t) + (\gamma_H + \delta_H), D(t)}{\alpha \delta_H + (\gamma_H + \delta_H)} \cdot \frac{1}{\delta_I} \right) \\
I'(t) &= k E(t) - (\alpha + \gamma_I + \delta_I) I(t), \\
H'(t) &= \alpha I(t) - (\gamma_H + \delta_H) H(t), \\
D'(t) &= \delta_I I(t) + \delta_H H(t), \\
x'(t) &= k E(t), \\
y(t) &= k E(t), \\
y_2(t) &= I(t) + (\alpha + \gamma_I + \delta_I) \cdot \left( \frac{\delta_H H(t) + (\gamma_H + \delta_H) D(t)}{\alpha \delta_H + (\gamma_H + \delta_H)} \cdot \frac{1}{\delta_I} \right) - C_3.
\end{align*}

The following is the reduced model used for structural identifiability analysis in JULIA. We refer to it as the reduced form of the original model.

\begin{equation}\label{Model7Red}\tag{M$_7$ Reduced}
\textbf{Model 7:}
\begin{cases}
 \displaystyle\frac{dE}{dt}=  \displaystyle\frac{(\beta_{I} I + \beta_{H} H+\beta_{D} D) (C_2 - E - (I +(\alpha + \gamma_I + \delta_I) \delta_H H + (\gamma_H+\delta_H)D)}{(C - D)(\alpha \delta_H + \gamma_H+\delta_H) \delta_I) - C_3))}-kE, \quad E(0) = E_0\\[1.5ex]
\displaystyle\frac{dI}{dt}=kE-(\alpha+\gamma_{I}+\delta_{I})I, \quad I(0) = I_0\\[1.5ex]
\displaystyle\frac{dH}{dt}=  \alpha I -(\gamma_{H}+\delta_{H}) H,\quad H(0) = H_0\\[1.5ex]  
\displaystyle\frac{dR}{dt}=  \gamma_I I +\gamma_{H} H, \quad R(0) = R_0\\[1.5ex]
 \displaystyle\frac{dD}{dt}=  \delta_{I} I+\delta_{H} H, \quad D(0) = D_0.\\[1.5ex]
 \end{cases}
\end{equation}

With this reduction, we perform identifiability analysis on the reduced model across three observation scenarios. \\ 

\textbf{(a) Number of new infected cases is observed}: For the first scenario, we assume the observations are the new infected cases. Julia results are shown in Table \ref{SI_M7a}. We summarize the findings in the following result.
\begin{table}[H]
\centering

\begin{minipage}[t]{0.47\linewidth}
\scriptsize
\textbf{(A)}\\[-0.3em]
\begin{verbatim}
ode = @ODEmodel(
  E'(t) = (betaI*I(t)+betaH*H(t)+betaD*D(t))*
  (C2-E(t)-(I(t)+(alpha+gammaI+deltaI)/
  (deltaH*alpha+(gammaH+deltaH)*deltaI)* 
  (deltaH*H(t)+(gammaH+deltaH)*D(t))-C3)) / 
  (C - D(t)) - k * E(t),
  I'(t) = k*E(t)-(alpha+gammaI+deltaI)*I(t),
  H'(t) = alpha*I(t)-(gammaH+deltaH)*H(t),
  D'(t) = deltaI*I(t)+deltaH*H(t),
  y(t) = k * E(t),
  y2(t) = I(t)+(alpha+gammaI+deltaI)/ 
  (deltaH*alpha+(gammaH+deltaH)*deltaI)* 
 (deltaH*H(t)+(gammaH+deltaH)*D(t))-C3
              )
julia> assess_identifiability(ode)

  E(t)   => :globally
  I(t)   => :locally
  H(t)   => :nonidentifiable
  D(t)   => :nonidentifiable
  C      => :nonidentifiable
  C2     => :globally
  C3     => :globally
  alpha  => :nonidentifiable
  betaD  => :globally
  betaH  => :nonidentifiable
  betaI  => :nonidentifiable
  deltaH => :nonidentifiable
  deltaI => :nonidentifiable
  gammaH => :nonidentifiable
  gammaI => :nonidentifiable
  k      => :globally

julia> find_identifiable_functions(ode)

 k
 betaD
 C3
 C2
 alpha + deltaH + deltaI + gammaH + gammaI
 alpha*deltaH + alpha*gammaH + deltaH*deltaI +
 deltaH*gammaI + deltaI*gammaH + gammaH*gammaI
 
 deltaI//betaI
 deltaI//C
 deltaI//(alpha*deltaH+deltaH*deltaI+deltaI
          *gammaH)
 (alpha*betaH+betaI*deltaH+betaI*gammaH)//
                                     deltaI

\end{verbatim}
\end{minipage}
\hspace{0.015\linewidth}
\vrule
\hspace{0.015\linewidth}
\begin{minipage}[t]{0.47\linewidth}
\scriptsize
\textbf{(B)}\\[-0.3em]
\begin{verbatim}
julia> assess_identifiability(ode, known_ic = [E,I,H,D])

  E(0)   => :globally
  I(0)   => :globally
  H(0)   => :globally
  D(0)   => :globally
  C      => :globally
  C2     => :globally
  C3     => :globally
  alpha  => :globally
  betaD  => :globally
  betaH  => :globally
  betaI  => :globally
  deltaH => :globally
  deltaI => :globally
  gammaH => :globally
  gammaI => :globally
  k      => :globally
\end{verbatim}
\end{minipage}

\caption{Structural identifiability analysis of Model~7a Reduced using the
\texttt{StructuralIdentifiability.jl} package in Julia.
Panel~(A) shows the model input--output formulation with unknown initial conditions,
while Panel~(B) shows the corresponding formulation with known initial conditions.}
\label{SI_M7a}
\end{table}

\begin{result}
    For the reduced model where the number of new infected cases is observed, structural identifiability analysis shows that when initial conditions are unknown, only the parameters $k$ and $\beta_D$ are globally structurally identifiable, while all other parameters remain unidentifiable. However, when initial conditions are known, all model parameters become structurally identifiable. This result underscores the importance of initial condition information and demonstrates that the reduced model is structurally identifiable under full observability of the initial state.
\end{result}

\textbf{(b) Number of new infections and hospitalizations are observed:} Next, we consider an augmented observation scenario by adding one additional observable—new hospitalizations—to the reduced system. In the model, new hospitalizations are given by $y_3(t) =\alpha I$. The corresponding input and output specifications for this setting are summarized in Table \ref{SI_M7b}. We summarize the identifiability analysis With this additional observation in the following:

\begin{table}[H]
\centering

\begin{minipage}[t]{0.47\linewidth}
\scriptsize
\textbf{(A)}\\[-0.3em]
\begin{verbatim}
ode = @ODEmodel(
 E'(t) = (betaI*I(t)+betaH*H(t)+betaD*D(t))
 *(C2-E(t)-(I(t)+(alpha+gammaI+deltaI)/
 (deltaH*alpha+(gammaH+deltaH)*deltaI)* 
 (deltaH*H(t)+(gammaH+deltaH)*D(t))-C3))/
 (C-D(t))-k*E(t),
 I'(t) = k*E(t)-(alpha+gammaI+deltaI)*I(t),
 H'(t) = alpha*I(t)-(gammaH+deltaH)*H(t),
 D'(t) = deltaI * I(t) + deltaH * H(t),
  y(t) = k * E(t),
  y2(t) = I(t)+(alpha+gammaI+deltaI)/ 
  (deltaH*alpha+(gammaH+deltaH)*deltaI)* 
  (deltaH*H(t)+(gammaH+deltaH)*D(t))-C3,
  y3(t) = alpha*I(t)
            )

julia> assess_identifiability(ode)

  E(t)   => :globally
  I(t)   => :globally
  H(t)   => :globally
  D(t)   => :nonidentifiable
  C      => :nonidentifiable
  C2     => :globally
  C3     => :globally
  alpha  => :globally
  betaD  => :globally
  betaH  => :nonidentifiable
  betaI  => :nonidentifiable
  deltaH => :nonidentifiable
  deltaI => :nonidentifiable
  gammaH => :nonidentifiable
  gammaI => :nonidentifiable
  k      => :globally

julia> find_identifiable_functions(ode)

 k
 betaD
 alpha
 C3
 C2
 deltaI + gammaI
 deltaH + gammaH
 deltaI//deltaH
 deltaI//betaI
 deltaI//betaH
 deltaI//C
            
\end{verbatim}
\end{minipage}
\hspace{0.015\linewidth}
\vrule
\hspace{0.015\linewidth}
\begin{minipage}[t]{0.47\linewidth}
\scriptsize
\textbf{(B)}\\[-0.3em]
\begin{verbatim}
julia> assess_identifiability(ode, known_ic = [E,I,H,D])

  E(0)   => :globally
  I(0)   => :globally
  H(0)   => :globally
  D(0)   => :globally
  C      => :globally
  C2     => :globally
  C3     => :globally
  alpha  => :globally
  betaD  => :globally
  betaH  => :globally
  betaI  => :globally
  deltaH => :globally
  deltaI => :globally
  gammaH => :globally
  gammaI => :globally
  k      => :globally
\end{verbatim}
\end{minipage}

\caption{Structural identifiability analysis of Model~7b Reduced using the
\texttt{StructuralIdentifiability.jl} package in Julia.
Panel~(A) shows the model input--output formulation with unknown initial conditions,
while Panel~(B) shows the corresponding formulation with known initial conditions.}
\label{SI_M7b}
\end{table}

\begin{pro}
      For the reduced model where both new infections and new hospitalizations are observed, structural identifiability analysis indicates that when initial conditions are unknown, only the parameters $k$, $\alpha$, and $\beta_D$ are globally structurally identifiable, while all other parameters remain unidentifiable. In contrast, when initial conditions are known, all parameters in the model become structurally identifiable. These findings demonstrate that incorporating a second observation significantly improves identifiability and that the reduced model is fully structurally identifiable under known initial conditions and dual observation streams.
\end{pro}
We see that adding new hospitalizations have only identified one more parameter, namely $\alpha$. Therefore to optian a structurally identifiable model, we need to add more observation. We do that next.

\textbf{(c) Number of new infections, hospitalizations and deaths are observed:} We now extend the analysis further by introducing a third observation—the number of disease-induced deaths—into the reduced system. Disease induced deaths are given in the model by $y_4(t) = \delta_I I +\delta_H H$

\begin{table}[H]
\centering

\begin{minipage}[t]{0.47\linewidth}
\scriptsize
\textbf{(A)}\\[-0.3em]
\begin{verbatim}
julia> ode = @ODEmodel(
    E'(t) = (betaI*I(t)+betaH*H(t)+betaD*
    D(t))*(C2-E(t)-(I(t)+(alpha+gammaI + 
    deltaI)/(deltaH*alpha+(gammaH+deltaH)*
    deltaI)*(deltaH*H(t)+(gammaH+deltaH)*
    D(t)) - C3)) / (C - D(t)) - k * E(t),
    I'(t) = k*E(t)-(alpha+gammaI+deltaI)*I(t),
    H'(t) = alpha*I(t)-(gammaH+deltaH)*H(t),
    D'(t) = deltaI*I(t)+deltaH*H(t),
     y(t) = k * E(t),
    y2(t) = I(t)+(alpha+gammaI+deltaI)/
    (deltaH*alpha+(gammaH+deltaH)*deltaI)*
    (deltaH*H(t)+(gammaH+deltaH)*D(t))-C3,
    y3(t) = alpha*I(t),
    y4(t) = deltaI*I(t)+deltaH*H(t)
            )

julia> assess_identifiability(ode)

  E(t)   => :globally
  I(t)   => :globally
  H(t)   => :globally
  D(t)   => :globally
  C      => :globally
  C2     => :globally
  C3     => :globally
  alpha  => :globally
  betaD  => :globally
  betaH  => :globally
  betaI  => :globally
  deltaH => :globally
  deltaI => :globally
  gammaH => :globally
  gammaI => :globally
  k      => :globally
\end{verbatim}
\end{minipage}
\hspace{0.015\linewidth}
\vrule
\hspace{0.015\linewidth}
\begin{minipage}[t]{0.47\linewidth}
\scriptsize
\textbf{(B)}\\[-0.3em]
\begin{verbatim}
julia> assess_identifiability(ode, known_ic = [E,I,H,D])

  E(0)   => :globally
  I(0)   => :globally
  H(0)   => :globally
  D(0)   => :globally
  C      => :globally
  C2     => :globally
  C3     => :globally
  alpha  => :globally
  betaD  => :globally
  betaH  => :globally
  betaI  => :globally
  deltaH => :globally
  deltaI => :globally
  gammaH => :globally
  gammaI => :globally
  k      => :globally
\end{verbatim}
\end{minipage}

\caption{Structural identifiability analysis of Model~7c Reduced using the
\texttt{StructuralIdentifiability.jl} package in Julia.
Panel~(A) shows the model input--output formulation with unknown initial conditions,
while Panel~(B) shows the corresponding formulation with known initial conditions.}
\label{SI_M7c}
\end{table}

With the inclusion of all three observations—new infections, hospitalizations, and deaths—the model achieves full structural identifiability regardless of whether the initial conditions are known.

\begin{pro}
    For the reduced model with three observational outputs, structural identifiability analysis shows that the model is globally structurally identifiable under both known and unknown initial conditions.
\end{pro}

This result clearly demonstrates that incorporating multiple independent observations can substantially improve parameter identifiability. In this case, the inclusion of three observation types ensures complete identifiability under all scenarios, reinforcing the principle that richer data streams are essential for robust model-based inference.

\subsection{COVID-19 Model}

The following model characterizes the transmission dynamics of respiratory infections such as COVID-19, where both pre-symptomatic and symptomatic individuals contribute to disease spread. The model comprises six epidemiological compartments: susceptible individuals $S(t)$, latent individuals $E(t)$, pre-symptomatic infectious individuals $I_{\rho}(t)$, symptomatic individuals $I(t)$, recovered individuals $R(t)$, and disease-induced deaths $D(t)$.

\begin{equation}\label{Model8}\tag{M$_8$}
\textbf{Model 8:}
\begin{cases}
\displaystyle\frac{dS}{dt} = -\frac{(\beta_{\rho} I_{\rho} + \beta_{I} I) S}{N}, \quad S(0) = S_0\\[1.5ex]
\displaystyle\frac{dE}{dt} = \frac{(\beta_{\rho} I_{\rho} + \beta_{I} I) S}{N}-kE, \quad E(0) = E_0\\[1.5ex]
\displaystyle\frac{dI_{\rho}}{dt}= kE - k_\rho I_{\rho} -  \gamma_{\rho} I_{\rho}, \quad I_{\rho}(0) = I_{\rho 0}\\[1.5ex]
\displaystyle\frac{dI}{dt} = k_\rho  I_{\rho}-\gamma I - \delta I, \quad I(0) = I_0\\[1.5ex]
\displaystyle\frac{dR}{dt} =  \gamma I +\gamma_{\rho} I_\rho, \quad R(0) = R_0\\[1.5ex]
\displaystyle\frac{dD}{dt} = \delta I, \quad D(0) = D_0.\\[1.5ex]
\end{cases}
\end{equation}

\begin{figure}[H]
    \centering
\includegraphics[scale=0.9]{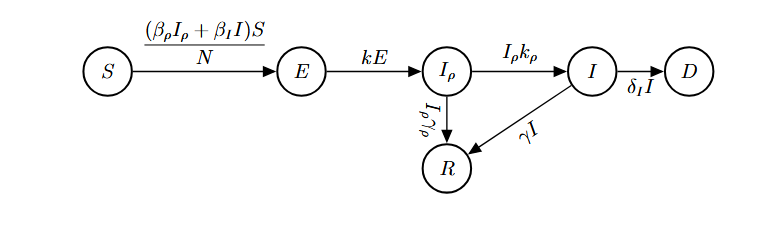}
    \caption{Flow diagram of Model \ref{Model8} showing the transmission dynamics of COVID-19. Diagram reproduced with permission from Chowell et al. (2023).}
    \label{fig:M8}
\end{figure}

We investigate the structural identifiability of Model~\ref{Model8} under two observational scenarios: (a) when only the number of new symptomatic cases is observed, and (b) when both new symptomatic cases and deaths are observed. \\ 

Due to the model complexity, direct application of structural identifiability tools to the original formulation of Model~\ref{Model8} under scenario (a) was not feasible. To overcome this limitation, we applied a model reduction strategy based on first integrals, as previously described, to eliminate the susceptible compartment and simplify the system. This allowed us to analyze the structural identifiability of the reduced model using available computational tools. 

The reduced model derived using the first integral method is expressed by the following system:

\begin{equation}\label{Model8Red}\tag{M$_8$ Reduced}
\textbf{Model 8:}
\begin{cases}
\displaystyle\frac{dE}{dt} = \frac{(\beta_{\rho} I_{\rho} + \beta_{I} I) ( c_1 - (E + I_{\rho} + (k_p +\gamma_p) \int y))}{N k_p}-kE, \quad E(0) = E_0\\[1.5ex]
\displaystyle\frac{dI_{\rho}}{dt}= kE - k_\rho I_{\rho} -  \gamma_{\rho} I_{\rho}, \quad I_{\rho}(0) = I_{\rho 0}\\[1.5ex]
\displaystyle\frac{dI}{dt} = k_\rho  I_{\rho}-\gamma I - \delta I, \quad I(0) = I_0\\[1.5ex]
\displaystyle\frac{dR}{dt} =  \gamma I +\gamma_{\rho} I_\rho, \quad R(0) = R_0\\[1.5ex]
\displaystyle\frac{dD}{dt} = \delta I, \quad D(0) = D_0.\\[1.5ex]
\end{cases}
\end{equation}

\textbf{(a) Number of new symptomatic cases is observed}.
We now perform the analysis for the case where the number of new symptomatic cases is observed using the reduced model. The model $M_8$ gives the new symptomatic cases as $y(t) =k_p I_p(t)$. The Julia input and output for the unknown and known initial condition scenarios are shown in Table~\ref{SI_M8a}. We summarize the findings from Table~\ref{SI_M8a} in the following result.

\begin{table}[H]
\centering

\begin{minipage}[t]{0.47\linewidth}
\scriptsize
\textbf{(A)}\\[-0.3em]
\begin{verbatim}
julia> ode= @ODEmodel(
 E'(t) = (beta_p*Ii(t)+beta_I*I(t))*(C1-(E+Ii+ 
 (k_p +gamma_p)*int_y/k_p))/(C-D(t)),
 Ii'(t) = k*E(t)-k_p*Ii(t)-gamma_p*Ii(t),
  I'(t) =  k_p*Ii(t) - gamma*I(t)-delta*I(t),
  D'(t) = delta*I(t),
  int_y'(t) = k_p*Ii(t),
  y(t) = k_p*Ii(t),
 y2(t) = int_y(t)
        )
julia>  assess_identifiability(ode)
  E(t)     => :nonidentifiable
  Ii(t)    => :nonidentifiable
  I(t)     => :globally
  D(t)     => :nonidentifiable
  int_y(t) => :globally
  C        => :nonidentifiable
  C1       => :nonidentifiable
  beta_I   => :nonidentifiable
  beta_p   => :nonidentifiable
  delta    => :nonidentifiable
  gamma    => :nonidentifiable
  gamma_p  => :nonidentifiable
  k        => :globally
  k_p      => :nonidentifiable

julia> find_identifiable_functions(ode)
 k
 C1*k_p
 gamma_p + k_p
 delta + gamma
 delta//beta_I
 delta//(C1*beta_p)




\end{verbatim}
\end{minipage}
\hspace{0.015\linewidth}
\vrule
\hspace{0.015\linewidth}
\begin{minipage}[t]{0.47\linewidth}
\scriptsize
\textbf{(B)}\\[-0.3em]
\begin{verbatim}
julia> assess_identifiability(ode, known_ic = [E,Ii,I,D])
  E(0)     => :globally
  Ii(0)    => :globally
  I(0)     => :globally
  D(0)     => :globally
  int_y(0) => :globally
  C        => :nonidentifiable
  C1       => :globally
  beta_I   => :nonidentifiable
  beta_p   => :nonidentifiable
  delta    => :nonidentifiable
  gamma    => :nonidentifiable
  gamma_p  => :globally
  k        => :globally
  k_p      => :globally
\end{verbatim}
\end{minipage}

\caption{Structural identifiability analysis of the reduced Model~8a using the
\texttt{StructuralIdentifiability.jl} package in Julia.
Panel~(A) shows the model input--output formulation with unknown initial conditions,
while Panel~(B) shows the corresponding formulation with known initial conditions.}
\label{SI_M8a}
\end{table}

\begin{result}
    The reduced model \ref{Model8} is not structurally identifiable when only the number of new symptomatic cases is observed. With unknown initial conditions, only the parameter $k$ is globally structurally identifiable. When initial conditions are known, the parameters $\gamma_{\rho}$, $k$, and $k_{\rho}$ become identifiable. Because several parameters remain unidentifiable in both scenarios, the model lacks full structural identifiability regardless of the assumptions made about initial conditions.
\end{result}

To address this limitation, we consider several strategies for improving identifiability. These include: (i) incorporating additional observational data, (ii) fixing certain parameters based on biological plausibility or prior information, and (iii) reformulating the model to reduce complexity or parameter redundancy. As a first step, we assess the impact of adding the number of new deaths as an additional observation.

It is also worth noting that the original unreduced version of the model~\ref{Model8} could be analyzed using \texttt{StructuralIdentifiability.jl} when two outputs were specified. We performed the identifiability analysis with the original unreduced version. In the following section, we summarize the identifiability results under that two-observation setting.

\textbf{(b) Number of new symptomatic cases  and deaths are observed.} We now extend the analysis to the case where both the number of new symptomatic cases and disease-induced deaths are observed. The Julia code and the resulting identifiability output for this scenario are shown in Table \ref{SI_M8b}.

According to the findings, the identifiability analysis incorporating the additional observations is summarized in the following result.

\begin{result}
   For the COVID-19 transmission model \ref{Model8}, when both the number of new symptomatic cases and deaths are observed: Under unknown initial conditions, the parameters $\beta_I$, $\delta$, and $\gamma$ are globally structurally identifiable, while $k$ is only locally identifiable. The parameters $\beta_{\rho}$, $\gamma_{\rho}$, and $k_{\rho}$ remain unidentifiable, likely due to insufficient variation in the observational outputs to disentangle their individual effects. In contrast, when the initial conditions are known, all model parameters become globally structurally identifiable. These findings underscore the importance of both informative data and accurate specification of initial conditions to achieve reliable parameter inference in complex transmission models.
\end{result}

\begin{table}[H]
\centering

\begin{minipage}[t]{0.47\linewidth}
\scriptsize
\textbf{(A)}\\[-0.3em]
\begin{verbatim}
julia> ode = @ODEmodel(
    S'(t) =  -(beta_p*Ii(t)+beta_I*I(t))*S(t)/
    (S(t)+E(t)+Ii(t)+I(t)+R(t)),
    E'(t) = (beta_p*Ii(t)+beta_I*I(t))*S(t)/
    (S(t)+E(t)+Ii(t)+I(t)+R(t))-(k*E(t)),
    Ii'(t) = k*E(t)-k_p*Ii(t)-gamma_p*Ii(t),
    I'(t) =  k_p*Ii(t)-gamma*I(t)-delta*I(t),
    R'(t) = gamma*I(t) + gamma_p*Ii(t),
    D'(t) = delta*I(t),
    y1(t) = k_p*Ii(t),
    y2(t) = delta*I(t)
        )

julia>  assess_identifiability(ode)
  S(t)    => :nonidentifiable
  E(t)    => :nonidentifiable
  Ii(t)   => :nonidentifiable
  I(t)    => :globally
  R(t)    => :nonidentifiable
  D(t)    => :nonidentifiable
  beta_I  => :globally
  beta_p  => :nonidentifiable
  delta   => :globally
  gamma   => :globally
  gamma_p => :nonidentifiable
  k       => :locally
  k_p     => :nonidentifiable

julia> find_identifiable_functions(ode)

 gamma
 delta
 beta_I
 gamma_p*k + k*k_p
 gamma_p + k + k_p
 k_p//beta_p
\end{verbatim}
\end{minipage}
\hspace{0.015\linewidth}
\vrule
\hspace{0.015\linewidth}
\begin{minipage}[t]{0.47\linewidth}
\scriptsize
\textbf{(B)}\\[-0.3em]
\begin{verbatim}
julia> assess_identifiability(ode, known_ic = [S,E,Ii,I,R,D])
  S(0)    => :globally
  E(0)    => :globally
  Ii(0)   => :globally
  I(0)    => :globally
  R(0)    => :globally
  D(0)    => :globally
  beta_I  => :globally
  beta_p  => :globally
  delta   => :globally
  gamma   => :globally
  gamma_p => :globally
  k       => :globally
  k_p     => :globally
\end{verbatim}
\end{minipage}

\caption{Structural identifiability analysis of Model~8b using the
\texttt{StructuralIdentifiability.jl} package in Julia.
Panel~(A) shows the model input--output formulation with unknown initial conditions,
while Panel~(B) shows the corresponding formulation with known initial conditions.}
\label{SI_M8b}
\end{table}


\subsection{SEUIR model}

Model \ref{Model9} describes a SEUIR compartmental model formulated as a system of ordinary differential equations. The state variables include the number of susceptible individuals $S(t)$, exposed individuals $E(t)$, symptomatic infectious individuals $I(t)$, unobserved or undocumented infectious individuals $U(t)$, and recovered individuals $R(t)$ at time $t$. This model structure allows for the explicit representation of both reported and unreported infectious individuals, which is especially relevant for diseases like COVID-19 with substantial underreporting. 

\begin{equation}\label{Model9}\tag{M$_9$}
\textbf{Model 9:}
\begin{cases}
    \displaystyle\frac{dS}{dt}&= \dfrac{-S(\beta_{I}I+\beta_{U}U)}{N}, \quad S(0) = S_0\\[1.5ex]
    \displaystyle\frac{dE}{dt}&= \dfrac{S(\beta_{I}I+\beta_{U}U)}{N}-(\kappa \rho+\kappa (1-\rho)) E, \quad E(0) = E_0\\[1.5ex]
    \displaystyle\frac{dI}{dt}&= \kappa \rho E-\gamma I,  \quad E(0) = E_0\\[1.5ex]
    \displaystyle\frac{dU}{dt}&= \kappa (1-\rho) E- \gamma U,  \quad I(0) = I_0\\[1.5ex]
    \displaystyle\frac{dR}{dt}&=\gamma I +\gamma U, \quad R(0) = R_0.\\[1.5ex]
\end{cases}
\end{equation}

The primary observational output is the number of new symptomatic cases, assumed to be proportional to $\kappa\rho E(t)$, where $\kappa$ is the progression rate from exposed to infectious and $\rho$ is the reporting fraction. The corresponding Julia code and the resulting identifiability output are presented in Table \ref{SI_M9}.

\begin{table}[H]
\centering

\begin{minipage}[t]{0.47\linewidth}
\scriptsize
\textbf{(A)}\\[-0.3em]
\begin{verbatim}
ode = @ODEmodel(
 S'(t) = -S(t)*(betai*I(t)+betau*U(t))/N,
 E'(t) = -(k*rho+k*(1-rho))*E(t)+S(t)*
 (betai*I(t)+betau*U(t))/N,
 I'(t) =k*rho*E(t)-gamma*I(t),
 U'(t)=k*(1-rho)*E(t)-gamma*U(t),
 R'(t)=gamma*I(t)+gamma*U(t),
 y1(t)=k*rho*E(t)
     )

julia> assess_identifiability(ode)

  S(t)  => :nonidentifiable
  E(t)  => :nonidentifiable
  I(t)  => :nonidentifiable
  U(t)  => :nonidentifiable
  R(t)  => :nonidentifiable
  N     => :nonidentifiable
  betai => :nonidentifiable
  betau => :nonidentifiable
  gamma => :globally
  k     => :globally
  rho   => :nonidentifiable

julia> find_identifiable_functions(ode)
 k
 gamma
 (betai*rho - betau*rho + betau)//(N*rho)
\end{verbatim}
\end{minipage}
\hspace{0.015\linewidth}
\vrule
\hspace{0.015\linewidth}
\begin{minipage}[t]{0.47\linewidth}
\scriptsize
\textbf{(B)}\\[-0.3em]
\begin{verbatim}
 ode = @ODEmodel(
    S'(t) = -S(t)*(betai*I(t)+betau*U(t))/N,
    E'(t) = -(k*rho+k*(1-rho))*E(t)+S(t)*(betai*I(t)+
    betau*U(t))/N,
    I'(t) =k*rho*E(t)-gamma*I(t),
     U'(t)=k*(1-rho)*E(t)-gamma*U(t),
     R'(t)=gamma*I(t)+gamma*U(t),
     y1(t)=k*rho*E(t),
     y2(t)=N
     )

 assess_identifiability(ode, known_ic = [S,E,I,U,R])

  S(0)  => :globally
  E(0)  => :globally
  I(0)  => :globally
  U(0)  => :globally
  R(0)  => :globally
  N     => :globally
  betai => :globally
  betau => :globally
  gamma => :globally
  k     => :globally
  rho   => :globally
\end{verbatim}
\end{minipage}

\caption{Structural identifiability analysis of Model~\ref{Model9} using the
\texttt{StructuralIdentifiability.jl} package in Julia.
Panel~(A) shows the model input--output formulation with unknown initial conditions,
while Panel~(B) shows the corresponding formulation with known initial conditions.}
\label{SI_M9}
\end{table}

\begin{figure}[H]
    \centering
\includegraphics[scale=0.8]{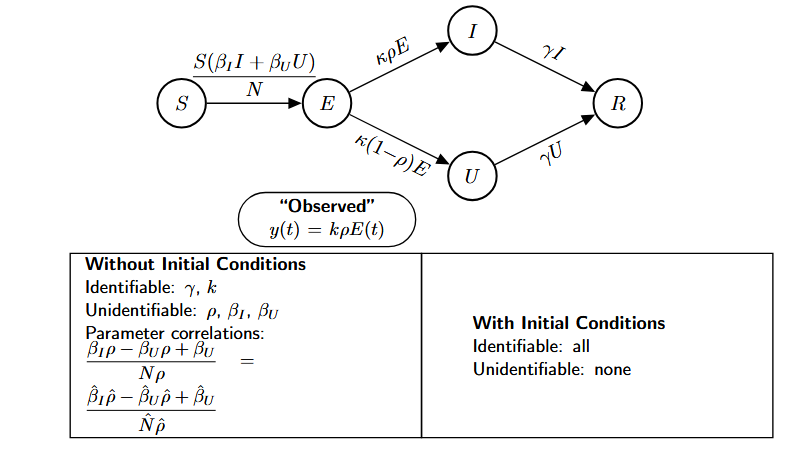}
    \caption{Flow diagram of the SEUIR model that accounts for both reported and unreported infectious individuals. The bottom panel presents structural identifiability results obtained using \texttt{StructuralIdentifiability.jl}, considering scenarios with both unknown and known initial conditions.}
    \label{fig:M9}
\end{figure}

\begin{result}
The SEUIR model \ref{Model9} is structurally unidentifiable when the initial conditions are unknown. Specifically, the parameters $\gamma$ and $k$ are globally identifiable, while the remaining parameters—including $\beta_I$, $\beta_U$, $\rho$, and $N$—cannot be uniquely determined from the observed data. In contrast, when the initial conditions are known, all model parameters become globally structurally identifiable.
\end{result}

\subsection{SEUIHRD model}

Model \ref{Model10} describes a SEUIHRD compartmental model formulated using a system of ordinary differential equations. The state variable $S(t)$ represents the number of susceptible individuals at time $t$, $E(t)$ denotes the number of exposed individuals, $I(t)$ corresponds to the number of symptomatic infected individuals, and $U(t)$ represents the number of unobserved or unaccounted infected individuals. $H(t)$ accounts for the number of hospitalized individuals, $D(t)$ denotes cumulative disease-induced deaths, and $R(t)$ is the number of recovered individuals. This model structure allows for a more realistic representation of infection progression, capturing heterogeneity in disease reporting, healthcare burden, and fatality.

\begin{equation}\label{Model10}\tag{M$_{10}$}
\textbf{Model 10:}
\begin{cases}
    \displaystyle\frac{dS}{dt}&= \dfrac{-S(\beta_{I}I+\beta_{U}U)}{N}, \quad S(0) = S_0\\[1.5ex]
    \displaystyle\frac{dE}{dt}&= \dfrac{S(\beta_{I}I+\beta_{U}U)}{N}-(\kappa \rho+\kappa (1-\rho)) E, \quad E(0) = E_0\\[1.5ex]
    \displaystyle\frac{dI}{dt}&= \kappa \rho E-(\gamma+\alpha) I , \quad I(0) = I_0\\[1.5ex]
    \displaystyle\frac{dU}{dt}&= \kappa (1-\rho) E- \gamma U, \quad U(0) = U_0\\[1.5ex]
    \displaystyle\frac{dH}{dt}&=\alpha I-\gamma_{H} H -\delta H, \quad H(0) = H_0\\[1.5ex]
    \displaystyle\frac{dD}{dt}&= \delta H, \quad D(0) = D_0\\[1.5ex] 
    \displaystyle\frac{dR}{dt}&=\gamma I +\gamma U+ \gamma_{H} H, \quad R(0) = R_0.\\[1.5ex] 
\end{cases}
\end{equation}

\begin{figure}[H]
    \centering
\includegraphics[scale=0.9]{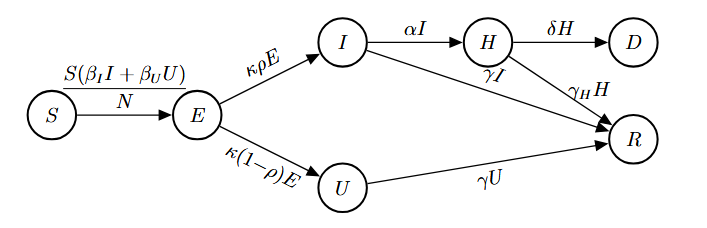}
    \caption{Flow diagram of the SEUIHRD model. This model captures key aspects of disease transmission and progression, including underreporting, hospitalization, and disease-induced mortality.}
    \label{fig:M10}
\end{figure}

In this model, we consider two observational scenarios:  
a) the observed quantities are the number of symptomatic cases, given by $\kappa\rho E$, and the number of hospitalizations, represented by $\alpha I$;  
b) the observed quantities include symptomatic cases ($\kappa\rho E$), hospitalizations ($\alpha I$), and deaths among hospitalized individuals, captured by $\gamma H$.  
These observational settings enable us to assess how the inclusion of additional health outcomes impacts the structural identifiability of model parameters. \\

\textbf{(a) Number of new reported infections and new hospitalizations are observed.}

\begin{table}[H]
\centering

\begin{minipage}[t]{0.47\linewidth}
\scriptsize
\textbf{(A)}\\[-0.3em]
\begin{verbatim}
ode = @ODEmodel(
    S'(t) = -S(t)*(betai*I(t)+betau*U(t))/N,
    E'(t) = -(k*rho+k*(1-rho))*E(t)+S(t)*
             (betai*I(t)+betau*U(t))/N,
    I'(t) =k*rho*E(t)-(gamma+alpha)*I(t),
    U'(t) =k*(1-rho)*E(t)-gamma*U(t),
    H'(t) =alpha*I(t)-gammah*H(t)-delta*H(t),
    D'(t) = delta*H(t),
    R'(t) =gamma*I(t)+gamma*U(t)+gammah*H(t),
    y1(t) =k*rho*E(t),
    y2(t) =alpha*I(t)
       )

julia> assess_identifiability(ode)

  S(t)   => :nonidentifiable
  E(t)   => :nonidentifiable
  I(t)   => :globally
  U(t)   => :nonidentifiable
  H(t)   => :nonidentifiable
  D(t)   => :nonidentifiable
  R(t)   => :nonidentifiable
  N      => :nonidentifiable
  alpha  => :globally
  betai  => :nonidentifiable
  betau  => :nonidentifiable
  delta  => :nonidentifiable
  gamma  => :globally
  gammah => :nonidentifiable
  k      => :globally
  rho    => :nonidentifiable

julia> find_identifiable_functions(ode)

 k
 gamma
 alpha
 betai//N
 (betau*rho - betau)//(betai*rho)
\end{verbatim}
\end{minipage}
\hspace{0.015\linewidth}
\vrule
\hspace{0.015\linewidth}
\begin{minipage}[t]{0.47\linewidth}
\scriptsize
\textbf{(B)}\\[-0.3em]
\begin{verbatim}
 ode = @ODEmodel(
    S'(t) = -S(t)*(betai*I(t)+betau*U(t))/N,
    E'(t) = -(k*rho+k*(1-rho))*E(t)+S(t)*(betai*I(t)
            +betau*U(t))/N,
    I'(t) =k*rho*E(t)-(gamma+alpha)*I(t),
    U'(t) =k*(1-rho)*E(t)-gamma*U(t),
    H'(t) =alpha*I(t)-gammah*H(t)-delta*H(t),
    D'(t) = delta*H(t),
    R'(t) =gamma*I(t)+gamma*U(t)+gammah*H(t),
    y1(t) =k*rho*E(t),
    y2(t) =alpha*I(t),
    y3(t) = N
        )

julia> assess_identifiability(ode, known_ic = [S,E,I,U,H,D,R])

  S(0)   => :globally
  E(0)   => :globally
  I(0)   => :globally
  U(0)   => :globally
  H(0)   => :globally
  D(0)   => :globally
  R(0)   => :globally
  N      => :globally
  alpha  => :globally
  betai  => :globally
  betau  => :globally
  delta  => :nonidentifiable
  gamma  => :globally
  gammah => :nonidentifiable
  k      => :globally
  rho    => :globally
\end{verbatim}
\end{minipage}

\caption{Structural identifiability analysis of Model~10a using the
\texttt{StructuralIdentifiability.jl} package in Julia.
Panel~(A) shows the model input--output formulation with unknown initial conditions,
while Panel~(B) shows the corresponding formulation with known initial conditions.}
\label{SI_M10a}
\end{table}

\begin{result}
The SEUIHRD model \ref{Model10} is not globally structurally identifiable when initial conditions are unknown. Specifically, the parameters $\alpha$, $\gamma$, and $k$ are globally identifiable, whereas the remaining parameters—including $\beta_I$, $\beta_U$, $\delta$, $\gamma_H$, $\rho$, and $N$—are not identifiable under this scenario. When all initial conditions are known, only $\delta$ and $\gamma_H$ remain unidentifiable. Thus, the model remains structurally unidentifiable whether or not the initial conditions are known. These findings underscore the need for additional observational data or model reformulation to enable full identifiability of the model parameters.
\end{result}

To improve the identifiability of the model, we extend the analysis by incorporating an additional observation—specifically, the number of new deaths. This allows us to assess the extent to which enriching the observational data improves parameter identifiability.\\

\textbf{(b) Number of new reported infections, new hospitalizations and new deaths are observed}

We now consider the scenario in which the number of new reported infections, new hospitalizations, and new deaths are observed. The Julia code and the resulting identifiability output for this case are shown in Table \ref{SI_M10b}.

\begin{table}[H]
\centering

\begin{minipage}[t]{0.47\linewidth}
\scriptsize
\textbf{(A)}\\[-0.3em]
\begin{verbatim}
julia>  ode = @ODEmodel(
    S'(t) = -S(t)*(betai*I(t)+betau*U(t))/N,
    E'(t) = -(k*rho+k*(1-rho))*E(t)+S(t)*
            (betai*I(t)+betau*U(t))/N,
    I'(t) =k*rho*E(t)-(gamma+alpha)*I(t),
    U'(t) =k*(1-rho)*E(t)-gamma*U(t),
    H'(t) =alpha*I(t)-gammah*H(t)-delta*H(t),
    D'(t) = delta*H(t),
    R'(t) =gamma*I(t)+gamma*U(t)+gammah*H(t),
    y1(t) =k*rho*E(t),
    y2(t) =alpha*I(t),
    y3(t) = delta*H(t)
       )

julia> assess_identifiability(ode)

  S(t)   => :nonidentifiable
  E(t)   => :nonidentifiable
  I(t)   => :globally
  U(t)   => :nonidentifiable
  H(t)   => :globally
  D(t)   => :nonidentifiable
  R(t)   => :nonidentifiable
  N      => :nonidentifiable
  alpha  => :globally
  betai  => :nonidentifiable
  betau  => :nonidentifiable
  delta  => :globally
  gamma  => :globally
  gammah => :globally
  k      => :globally
  rho    => :nonidentifiable

julia> find_identifiable_functions(ode)

 k
 gammah
 gamma
 delta
 alpha
 betai//N
 (betau*rho - betau)//(betai*rho)
\end{verbatim}
\end{minipage}
\hspace{0.015\linewidth}
\vrule
\hspace{0.015\linewidth}
\begin{minipage}[t]{0.47\linewidth}
\scriptsize
\textbf{(B)}\\[-0.3em]
\begin{verbatim}

julia>  ode = @ODEmodel(
    S'(t) = -S(t)*(betai*I(t)+betau*U(t))/N,
    E'(t) = -(k*rho+k*(1-rho))*E(t)+S(t)*(betai*I(t)+
            betau*U(t))/N,
    I'(t) =k*rho*E(t)-(gamma+alpha)*I(t),
    U'(t) =k*(1-rho)*E(t)-gamma*U(t),
    H'(t) =alpha*I(t)-gammah*H(t)-delta*H(t),
    D'(t) = delta*H(t),
    R'(t) =gamma*I(t)+gamma*U(t)+gammah*H(t),
    y1(t) =k*rho*E(t),
    y2(t) =alpha*I(t),
    y3(t) = delta*H(t),
    y4(t) = N
    )

julia> assess_identifiability(ode, known_ic = [S,E,I,U,H,D,R])

  S(0)   => :globally
  E(0)   => :globally
  I(0)   => :globally
  U(0)   => :globally
  H(0)   => :globally
  D(0)   => :globally
  R(0)   => :globally
  N      => :globally
  alpha  => :globally
  betai  => :globally
  betau  => :globally
  delta  => :globally
  gamma  => :globally
  gammah => :globally
  k      => :globally
  rho    => :globally
\end{verbatim}
\end{minipage}

\caption{Structural identifiability analysis of Model~10b using the
\texttt{StructuralIdentifiability.jl} package in Julia.
Panel~(A) shows the model input--output formulation with unknown initial conditions,
while Panel~(B) shows the corresponding formulation with known initial conditions.}
\label{SI_M10b}
\end{table}

We summarize the findings from this extended analysis in the following result.

\begin{result}
   For the SEUIHRD model \ref{Model10}, when three types of observations are available—new reported infections, new hospitalizations, and new deaths—the following results hold: With unknown initial conditions, the parameters $\alpha$, $\delta$, $\gamma$, $\gamma_H$, and $k$ are globally structurally identifiable, while the remaining parameters remain unidentifiable. When the initial conditions are known, the model becomes fully identifiable, and all parameters can be structurally identified. These results highlight the critical role of incorporating multiple observations to resolve identifiability issues in complex compartmental models.
\end{result}

\subsection{SEIR model with spillover infections and human-to-human transmission }

Model \ref{Model11} captures the dynamics of spillover infections from poultry and subsequent human-to-human transmission. The model includes six compartments: $S(t)$ represents the number of susceptible humans; $E_{i}(t)$ denotes the number of exposed humans infected through interspecies (spillover) transmission; and $E_{s}(t)$ represents the number of exposed humans infected through secondary (human-to-human) transmission. $I_{i}(t)$ refers to infectious humans arising from spillover infection, $I_{s}(t)$ denotes infectious humans infected via human-to-human transmission, and $R(t)$ represents recovered humans. The parameter $\alpha$ denotes the per-capita spillover force of infection due to contact with infectious poultry and acts as an exogenous transmission pressure from the reservoir. The analysis is conducted over time horizons for which all state variables remain non-negative. This model structure enables the characterization of both zoonotic spillover events and sustained human transmission chains.

\begin{equation}\label{Model11}\tag{M$_{11}$}
\textbf{Model 11:}
\begin{cases}
    \dfrac{dS}{dt}&= \dfrac{-\beta S(I_{i}+I_{s})}{N}-\alpha, \quad S(0) = S_0\\[1.5ex]
    \dfrac{dE_{i}}{dt}&=\alpha-\kappa E_{i}, \quad E_i(0) = E_{i_0}\\[1.5ex]
    \dfrac{dI_{i}}{dt}&=\kappa E_{i}-\gamma I_{i}, \quad I_i(0) = I_{i_0}\\[1.5ex]
    \dfrac{dE_{s}}{dt}&=\dfrac{\beta S(I_{i}+I_{s})}{N}-\kappa E_{s}, \quad E_s(0) = E_{s_0}\\[1.5ex]
    \dfrac{dI_{s}}{dt}&=\kappa E_{s}-\gamma I_{s} \quad I_s(0) = I_{s_0}\\[1.5ex]
    \dfrac{dR}{dt}&=\gamma(I_{i}+I_{s}), \quad R(0) = R_0.\\[1.5ex]
\end{cases}
\end{equation}

The observable measures will be the spillover cases ($\kappa E_{i}$) and human-to-human cases ($\kappa E_{s}$). 

\begin{table}[H]
\centering

\begin{minipage}[t]{0.47\linewidth}
\scriptsize
\textbf{(A)}\\[-0.3em]
\begin{verbatim}
ode = @ODEmodel(
              S'(t) = -alpha-beta*(Ii+Is)/N,
              Ei'(t) = alpha - k*Ei,
              Ii'(t) = k*Ei - gamma*Ii,
              Es'(t) = -k*Es + beta*(Ii+Is)/N,
              Is'(t) = k*Es -gamma*Is,
              R'(t) = gamma*(Ii+Is),
              y1(t) = k*Ei,
              y2(t) = k*Es
              )

julia> assess_identifiability(ode)

   S(t)  => :nonidentifiable
  Ei(t) => :globally
  Ii(t) => :nonidentifiable
  Es(t) => :globally
  Is(t) => :nonidentifiable
  R(t)  => :nonidentifiable
  N     => :nonidentifiable
  alpha => :globally
  beta  => :nonidentifiable
  gamma => :globally
  k     => :globally

julia> find_identifiable_functions(ode)

  k
 gamma
 alpha
 beta//N
\end{verbatim}
\end{minipage}
\hspace{0.015\linewidth}
\vrule
\hspace{0.015\linewidth}
\begin{minipage}[t]{0.47\linewidth}
\scriptsize
\textbf{(B)}\\[-0.3em]
\begin{verbatim}
ode = @ODEmodel(
              S'(t) = -alpha-beta*(Ii+Is)/N,
              Ei'(t) = alpha - k*Ei,
              Ii'(t) = k*Ei - gamma*Ii,
              Es'(t) = -k*Es + beta*(Ii+Is)/N,
              Is'(t) = k*Es -gamma*Is,
              R'(t) = gamma*(Ii+Is),
              y1(t) = k*Ei,
              y2(t) = k*Es,
              y3(t) = N

julia> assess_identifiability(ode, known_ic = [S,Ei,Ii,Es,Is,R])

  S(0)  => :globally
  Ei(0) => :globally
  Ii(0) => :globally
  Es(0) => :globally
  Is(0) => :globally
  R(0)  => :globally
  N     => :globally
  alpha => :globally
  beta  => :globally
  gamma => :globally
  k     => :globally
\end{verbatim}
\end{minipage}

\caption{Structural identifiability analysis of Model~11 using the
\texttt{StructuralIdentifiability.jl} package in Julia.
Panel~(A) shows the model input--output formulation with unknown initial conditions,
while Panel~(B) shows the corresponding formulation with known initial conditions.}
\label{SI_M11}
\end{table}

\begin{figure}[H]
    \centering
\includegraphics[scale=0.7]{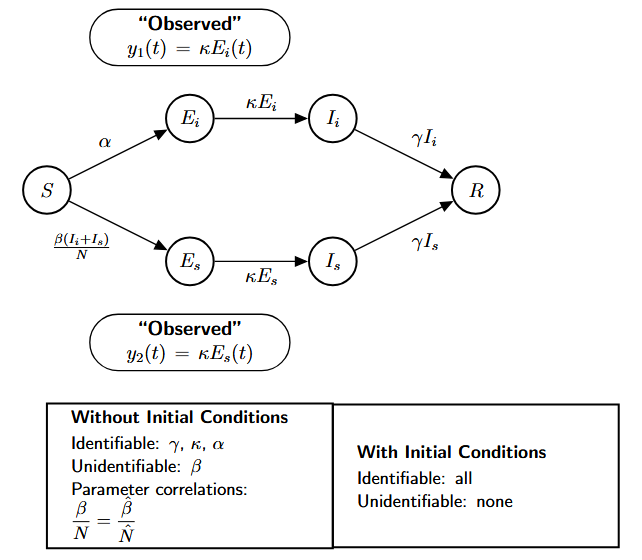}
    \caption{Flow diagram of the SEIR model incorporating both zoonotic spillover infections and human-to-human transmission pathways. The bottom panel presents structural identifiability results obtained using \texttt{StructuralIdentifiability.jl}, considering scenarios with both unknown and known initial conditions.}
    \label{fig:M11}
\end{figure}

\begin{result}
   The SEIR model with spillover infections and human-to-human transmission is not globally structurally identifiable when the initial conditions are unknown. Specifically, the parameters $k$, $\alpha$, and $\gamma$ are globally identifiable, whereas $\beta$ and $N$ are not structurally identifiable. However, when the initial conditions are known, all model parameters become globally structurally identifiable.
\end{result}

We now summarize the structural identifiability findings for all models using both the \texttt{StructuralIdentifiability.jl} package in Julia and the DAISY software. The results for Models \ref{Model1}--\ref{Model8} from DAISY were originally presented in our previous work \cite{Chowell2023}. Across the models analyzed, we found strong agreement between the results produced by \texttt{StructuralIdentifiability.jl} and DAISY. Notably, DAISY did not yield results for Models 7a and 7b, as indicated in \cite{Chowell2023}, nor for their reduced forms. Similarly, DAISY was unable to process Model 8a or its reduced version. These computational limitations underscore the utility of Julia-based symbolic tools such as \texttt{StructuralIdentifiability.jl} for analyzing more complex epidemic models.\\

Table \ref{Softwares} presents the identifiability results under unknown initial conditions, while Table \ref{Softwares_IC} summarizes the results when initial conditions are assumed to be known.

To further evaluate the identifiability of the models and the different methods, we extended the comparison to additional widely used tools, namely COMBOS (the Identifiable Combinations web application)\cite{combos}, SIAN (Structural Identifiability ANalyzer) \cite{SIAN}, and STRIKE-GOLDD (STRuctural Identifiability taKen as Extended-Generalized Observability with Lie Derivatives and Decomposition)\cite{Strike}. All analyses were conducted using the same observations and assumptions to ensure comparability.

COMBOS is based on an input–output equation approach and assesses global structural identifiability through symbolic solution of algebraic systems \cite{combos,Hong2020}. SIAN uses symbolic computation and differential algebra to evaluate both local and global structural identifiability of nonlinear dynamical systems \cite{SIAN,Hong2020}. STRIKE-GOLDD relies on differential-geometric methods, employing Lie derivatives and generalized observability analysis to assess local structural identifiability \cite{Strike,Strike2}.

SIAN and COMBOS provide both local and global structural identifiability analyses, whereas STRIKE-GOLDD is limited to local identifiability. In addition, SIAN and COMBOS identify identifiable parameter correlations or combinations, while COMBOS and STRIKE-GOLDD allow identifiability to be assessed under both known and unknown initial condition assumptions.

The results obtained with SIAN were generally consistent with those obtained using \texttt{StructuralIdentifiability.jl} in JULIA (see Table \ref{Softwares_SIAN}). In most cases, both tools produced the same identifiability conclusions. However, in some scenarios, SIAN was unable to assess global identifiability and returned only local identifiability results. These differences do not reflect substantive disagreement between the methods, but rather differences in methodological scope, particularly with respect to the distinction between local and global structural identifiability. STRIKE-GOLDD provided local identifiability results under both known and unknown initial condition assumptions, but for some models it did not return any results (see Table \ref{Softwares_Strike_GOLDD}). COMBOS did not provide identifiability results for any of the considered models, including simple models. Taken together with the results reported in Tables \ref{Softwares} and \ref{Softwares_IC}, these findings indicate that structural non-identifiability in many epidemic models arises primarily from unknown initial conditions rather than from model complexity per se, and that incorporating prior information on initial states can restore full identifiability without altering model structure.

We used the SIAN Maple Cloud application \cite{SIAN}, the COMBOS web application \cite{combos}, and STRIKE-GOLDD implemented in MATLAB R2024b \cite{Strike}.

\newpage

\begin{table}[H]
	
 \footnotesize
	\hspace*{-3.5cm}
		\footnotesize
\begin{tabular}{|*{6}{c|}}
\hline
 & & \multicolumn{2}{c|}{DAISY} &\multicolumn{2}{c|}{\texttt{\texttt{StructuralIdentifiability.jl}} in JULIA}\\

\hline
Model& Observations   & \begin{tabular}{c}  \textbf{Identifiable}\\ \textbf{parameters}  \end{tabular} &\begin{tabular}{c}  \textbf{Unidentifiable} \\ \textbf{parameters}\\  \end{tabular}& \begin{tabular}{c} \textbf{Identifiable}\\ \textbf{parameters}  \end{tabular} & \begin{tabular}{c}  \textbf{Unidentifiable} \\ \textbf{parameters} \\ \end{tabular}\\
\hline
M1 & \begin{tabular}{c} Number of new \\infected cases $(kE)$  \end{tabular}& $k$, $\gamma$ & $\beta$, $N$ & $k$, $\gamma$ & $\beta$, $N$\\
\hline
M2 &  \begin{tabular}{c} Number of new\\ infected sympt-\\omatic cases $(k \rho E)$\end{tabular} &$k$, $\gamma$ & $\rho$, $\beta$, $N$ & $k$, $\gamma$ & $\rho$, $\beta$, $N$\\
\hline
M3 &\begin{tabular}{c} Number of new\\ infected sympt-\\omatic cases $(k \rho E)$ \end{tabular} &$k$, $\gamma$ & $\rho$, $\beta_A$, $\beta_I$, $N$ & $k$, $\gamma$ & $\rho$, $\beta_A$, $\beta_I$, $N$ \\
\hline 
M4a & \begin{tabular}{c} Number of new \\infected cases $(kE)$ \end{tabular} &$k$ & $\beta,\delta, \gamma$ & $k$ & $\beta,\delta, \gamma$\\
\hline
M4b &\begin{tabular}{c} Number of new \\infected cases $(kE)$ \\and new deaths $(\delta I)$  \end{tabular} &$k$, $\beta$, $\delta$, $\gamma$  & - & $k$, $\beta$, $\delta$, $\gamma$& -\\
\hline
M5 &\begin{tabular}{c} Cumulative number \\of new incidence \\$(S(0)- S(t))$ \end{tabular} &$\gamma$, $\mu_\upsilon$ & $\beta$, $\beta_\upsilon$, $\Lambda_\upsilon$, $N$  & $\gamma$, $\mu_\upsilon$ & $\beta$, $\beta_\upsilon$, $\Lambda_\upsilon$, $N$\\
\hline
M6 &\begin{tabular}{c} Cumulative number \\of new incidence \\$(S(0)- S(t))$ \end{tabular} &$\gamma$,$\mu_\upsilon$ & $\beta_A$, $\beta_I$, $\beta_\upsilon$, $\rho$, $\Lambda_\upsilon$, $N$ & $\gamma$,$\mu_\upsilon$ & $\beta_A$, $\beta_I$, $\beta_\upsilon$, $\rho$, $\Lambda_\upsilon$, $N$ \\
\hline
\begin{tabular}{c}M7a\\ (reduced) \end{tabular}& \begin{tabular}{c} Number of new \\infected cases $(kE)$\end{tabular}& no results & no results & $k$,$\beta_D$& \begin{tabular}{c}$\beta_I,\beta_H,\alpha,\delta_I,$\\$\delta_H,\gamma_H,\gamma_I$ \end{tabular}\\
\hline
\begin{tabular}{c}M7b\\ (reduced) \end{tabular}&\begin{tabular}{c} Number of new \\infected cases $(kE)$ \\and new
hospitalization $(\alpha I)$ \end{tabular} &no results & no results&$\alpha$, $\beta_D$, $k$& $\beta_I,\beta_H,\delta_I,\delta_H,\gamma_H,\gamma_I$\\
\hline
\begin{tabular}{c}M7c\\ (reduced) \end{tabular} &\begin{tabular}{c} Number of new infected\\ cases $(kE)$, new \\hospitalization
$(\alpha I)$ and, \\new deaths $(\delta_I I + \delta_H H)$  \end{tabular} &all& - &all& -\\
\hline
\begin{tabular}{c}M8a\\ (reduced) \end{tabular} &\begin{tabular}{c} Number of new \\symptomatic cases $(k_{\rho} I_{\rho})$  \end{tabular} &no results& no results&$k$& $\gamma, \delta, \gamma_{\rho}, k_\rho, \beta_{\rho}, \beta_I$ \\
\hline
M8b & \begin{tabular}{c} Number of new symptomatic \\cases $(k_{\rho} I_{\rho})$ and new deaths \\$(\delta_I I)$  \end{tabular} &$\gamma, \delta, k, \beta_I$ & $\beta_\rho, \gamma_\rho, k_\rho$ &$\gamma, \delta, k\textit{(locally)}, \beta_I$ & $\beta_\rho, \gamma_\rho, k_\rho$\\
\hline
M9 &\begin{tabular}{c} New reported cases ($k \rho E$)  \end{tabular} & $\gamma, k $& $\beta_I, \beta_U, \rho$ & $\gamma, k $& $\beta_I, \beta_U, \rho$ \\
\hline
M10a & \begin{tabular}{c} new reported cases \\($k \rho E$), new \\hospitalizations ($\alpha I$)\end{tabular} & $\alpha, \gamma, k$ &$ \beta_I,\beta_U, \delta,\gamma_H,\rho$ & $\alpha,  \gamma, k$ &$\beta_I,\beta_U, \delta,\gamma_H,\rho$ \\
\hline
M10b &\begin{tabular}{c}new reported cases\\ ($k \rho E$), new \\hospitalizations ($\alpha I$), \\ new deaths ($\delta H$) \end{tabular} &$\alpha, \delta,\gamma,\gamma_H,k$ & $\beta_I,\beta_U,\rho$&$\alpha, \delta,\gamma,\gamma_H,k$ & $\beta_I,\beta_U,\rho$\\
\hline
M11 &\begin{tabular}{c} new spillover cases ($k E_i$),\\ new human-to-human \\cases ($k E_s$)\end{tabular} & $k, \alpha, \gamma$ &all &$k, \alpha, \gamma$ &all\\
\hline
\end{tabular} 
\caption{Summary of structural identifiability results for all models under unknown initial conditions, as obtained using the DAISY software and the \texttt{StructuralIdentifiability.jl} package in Julia. This table highlights the prevalence of structural non-identifiability under unknown initial conditions, even in widely used epidemic models. Entries labeled ‘no results’ indicate cases where the corresponding software tool was unable to complete the analysis within practical computational limits.}
	 \label{Softwares}
\end{table}

\begin{table} [H]
	
 \footnotesize
	\hspace*{-3cm}
    \footnotesize
\begin{tabular}{|*{6}{c|}}
\hline
 & & \multicolumn{2}{c|}{DAISY} &\multicolumn{2}{c|}{\texttt{\texttt{StructuralIdentifiability.jl}}}\\

\hline
Model& Observations   & \begin{tabular}{c}  \textbf{Identifiable}\\ \textbf{parameters}  \end{tabular} &\begin{tabular}{c}  \textbf{Unidentifiable} \\ \textbf{parameters}\\  \end{tabular}& \begin{tabular}{c} \textbf{Identifiable}\\ \textbf{parameters}  \end{tabular} & \begin{tabular}{c}  \textbf{Unidentifiable} \\ \textbf{parameters} \\ \end{tabular}\\
\hline
M1 &  \begin{tabular}{c} Number of new \\infected cases $(kE)$  \end{tabular}&  all &-& all &-\\
\hline
M2 &\begin{tabular}{c} Number of new\\ infected sympt-\\omatic cases $(k \rho E)$\end{tabular} & all& - & all& -\\
\hline
M3 &\begin{tabular}{c} Number of new\\ infected sympt-\\omatic cases $(k \rho E)$ \end{tabular} & all & - &all & -  \\
\hline 
M4a  & \begin{tabular}{c} Number of new \\infected cases $(kE)$ \end{tabular} & all & - & all & -\\
\hline
M4b & \begin{tabular}{c} Number of new \\infected cases $(kE)$ \\and new deaths $(\delta I)$  \end{tabular} &all  & - &  all  & -\\
\hline
M5 & \begin{tabular}{c} Cumulative number \\of new incidence \\$(S(0)- S(t))$ \end{tabular} &all & -  & all & -\\
\hline
M6 & \begin{tabular}{c} Cumulative number \\of new incidence \\$(S(0)- S(t))$ \end{tabular} & all & - &  all & -\\
\hline
\begin{tabular}{c}M7a\\ (reduced) \end{tabular} & \begin{tabular}{c} Number of new \\infected cases $(kE)$\end{tabular}& no results &  no results & all & -\\
\hline
\begin{tabular}{c}M7b\\ (reduced) \end{tabular}&\begin{tabular}{c} Number of new \\infected cases $(kE)$ \\and new\\
hospitalization $(\alpha I)$ \end{tabular}  &no results & no results& all & -\\
\hline
\begin{tabular}{c}M7c\\ (reduced) \end{tabular} &\begin{tabular}{c} Number of new infected\\ cases $(kE)$, new \\hospitalization
$(\alpha I)$ and, \\new deaths $(\delta_I I + \delta_H H)$  \end{tabular} & all & -&all& -\\
\hline
M8a & \begin{tabular}{c} Number of new \\symptomatic cases $(k_{\rho} I_{\rho})$ \end{tabular} &no results& no results&$k, \gamma_{\rho}, k_{\rho}$& $\gamma, \delta, \beta_{\rho}, \beta_I$ \\
\hline
M8b & \begin{tabular}{c} Number of new \\symptomatic cases $(k_{\rho} I_{\rho})$ \\and new deaths $(\delta_I I)$  \end{tabular} &all & - & all & -\\
\hline
M9 &\begin{tabular}{c} New reported cases ($k \rho E$)  \end{tabular} &all& -&all& -  \\
\hline
M10a & \begin{tabular}{c} new reported cases \\($k \rho E$), new \\hospitalizations ($\alpha I$)\end{tabular} & $\alpha, \beta_I, \beta_U,\gamma,k,\rho$ &$\delta, \gamma_H$& $\alpha, \beta_I, \beta_U,\gamma,k,\rho$ &$\delta, \gamma_H$  \\
\hline
M10b &\begin{tabular}{c}new reported cases\\ ($k \rho E$), new \\hospitalizations ($\alpha I$), \\ new deaths ($\delta H$) \end{tabular} &all& -&all& -  \\
\hline
M11 &\begin{tabular}{c} new spillover cases ($k E_i$),\\ new human-to-human \\cases ($k E_s$)\end{tabular} &all& -&all& -  \\
\hline
\end{tabular}
\caption{Summary of structural identifiability results for all models under known initial conditions, as obtained using the DAISY software and the \texttt{StructuralIdentifiability.jl} package in Julia. Entries labeled ‘no results’ indicate cases where the corresponding software tool was unable to complete the analysis within practical computational limits.}
	 \label{Softwares_IC}
\end{table}

\begin{table}[H]
\centering
\begin{tabular}{|*{3}{c|}}
\hline
 & \multicolumn{2}{c|}{SIAN} \\
\hline
Model   & \begin{tabular}{c}  \textbf{Identifiable}\\ \textbf{parameters}  \end{tabular} 
& \begin{tabular}{c}  \textbf{Unidentifiable} \\ \textbf{parameters}\\  \end{tabular} \\
\hline
M1& $k$, $\gamma$ & $\beta$, $N$ \\
\hline
M2 &$k$, $\gamma$ & $\rho$, $\beta$, $N$ \\
\hline
M3  &$k$, $\gamma$ & $\rho$, $\beta_A$, $\beta_I$, $N$ \\
\hline 
M4a &$k$ & $\beta,\delta, \gamma$ \\
\hline
M4b &$k$, $\beta$, $\delta$, $\gamma$  & - \\
\hline
M5  &$\gamma$, $\mu_\upsilon$ & $\beta$, $\beta_\upsilon$, $\Lambda_\upsilon$, $N$ \\
\hline
M6  &$^{*}\gamma$,$\mu_\upsilon$ & $\beta_A$, $\beta_I$, $\beta_\upsilon$, $\Lambda_\upsilon$, $N$,$\rho$ \\
\hline
\begin{tabular}{c}M7a\\ (reduced) \end{tabular}
& $^{*}k$,$\beta_D$ 
& \begin{tabular}{c}$\beta_I,\beta_H,\alpha,\delta_I,$\\$\delta_H,\gamma_H,\gamma_I$ \end{tabular} \\
\hline
\begin{tabular}{c}M7b\\ (reduced) \end{tabular}
& $^{*}\alpha$,$\beta_D$,$k$ 
& $\beta_I,\beta_H,\delta_I,\delta_H,\gamma_H,\gamma_I$ \\
\hline
\begin{tabular}{c}M7c\\ (reduced) \end{tabular} 
& all & - \\
\hline
\begin{tabular}{c}M8a\\ (reduced) \end{tabular} 
& $^{*}k$
& $\gamma, \delta, \gamma_{\rho}, k_\rho, \beta_{\rho}, \beta_I$ \\
\hline
M8b 
& $\gamma, \delta, k \textit{(locally)}, \beta_I$ 
& $\beta_\rho, \gamma_\rho, k_\rho$ \\
\hline
M9  
& $\gamma, k $ 
& $\beta_I, \beta_U, \rho$ \\
\hline
M10a 
& $\alpha,  \gamma, k$ 
& $N,\beta_U, \delta,\gamma_H,\rho, \beta_I$ \\
\hline
M10b 
& $\alpha, \delta,\gamma,\gamma_H,k$ 
& $N,\beta_U,\rho,\beta_I$ \\
\hline
M11 
& $k, \alpha, \gamma$ 
& $N, \beta$ \\
\hline
\end{tabular} 

\caption{Structural identifiability results obtained using SIAN for all considered models (M1–M11). Identifiable and unidentifiable parameters are reported for each model. The symbol $*$ indicates cases in which SIAN was unable to assess global identifiability and returned only local identifiability results.}
\label{Softwares_SIAN}
\end{table}

\begin{table}[H]
\centering
	 \begin{tabular}{|*{5}{c|}}
\hline
 \multicolumn{5}{|c|}{Strike-GOLDD: Locally structural identifiability} \\
\hline
 & \multicolumn{2}{c|}{With unknown ICs} &\multicolumn{2}{c|}{With known ICs}\\
\hline
Model   & \begin{tabular}{c}  \textbf{Identifiable}\\ \textbf{parameters}  \end{tabular} &\begin{tabular}{c}  \textbf{Unidentifiable} \\ \textbf{parameters}\\  \end{tabular}& \begin{tabular}{c} \textbf{Identifiable}\\ \textbf{parameters}  \end{tabular} & \begin{tabular}{c}  \textbf{Unidentifiable} \\ \textbf{parameters} \\ \end{tabular}\\
\hline
M1& $k$, $\gamma$ & $\beta$, $N$ & all & -\\
\hline
M2 &$k$, $\gamma$ & $\rho$, $\beta$, $N$ & $k$, $\gamma$ , $\beta$, $N$& $\rho$\\
\hline
M3  &no results& no results & $k$, $\gamma$,$\rho$,$N$ &  $\beta_A$, $\beta_I$ \\
\hline 
M4a &no results & no results & no results & no results\\
\hline
M4b &all  & - & all& -\\
\hline
M5  &$\gamma$, $\mu_\upsilon$ & $\beta$, $\beta_\upsilon$, $\Lambda_\upsilon$, $N$  & $\gamma$, $\mu_\upsilon$ & $\beta$, $\beta_\upsilon$, $\Lambda_\upsilon$, $N$\\
\hline
M6  &no results& no results & $\gamma$,$\mu_\upsilon$,$\rho$ & $\beta_A$, $\beta_I$, $\beta_\upsilon$, $\Lambda_\upsilon$, $N$ \\
\hline
\begin{tabular}{c}M7a\\ (reduced) \end{tabular}& no results & no results & no results& no results\\
\hline
\begin{tabular}{c}M7b\\ (reduced) \end{tabular}&no results&no results& all&-\\
\hline
\begin{tabular}{c}M7c\\ (reduced) \end{tabular} &all& - &all& -\\
\hline
\begin{tabular}{c}M8a\\ (reduced) \end{tabular}& no results & no results & no results& no results\\
\hline
M8b &  no results & no results & no results& no results\\
\hline
M9  &  no results&  no results& $N,\gamma, k, \rho $& $\beta_I, \beta_U$ \\
\hline
M10a &$N,\alpha,  \gamma, k$ &$\beta_U, \delta,\gamma_H,\rho, \beta_I,$ &$N$, $\alpha, \beta_I, \gamma, k$ &$\beta_U, \delta,\gamma_H,\rho$ \\
\hline
M10b &$N,\alpha, \delta,\gamma,\gamma_H,k$ & $\beta_U,\rho,\beta_I$&$N,\alpha, \beta_I,\delta,\gamma,\gamma_H,k,$ & $\beta_U,\rho$\\
\hline
M11 & $k, \alpha, \gamma$ &$N, \beta$ &all &-\\
\hline
\end{tabular} 
\caption{Local structural identifiability results obtained using STRIKE-GOLDD for all considered models (M1–M11). For each model, identifiable and unidentifiable parameters are reported under two scenarios: unknown initial conditions (ICs) and known initial conditions. The entry “all” indicates that all model parameters are identifiable, while “–” denotes that no unidentifiable parameters remain. “No results” indicates that STRIKE-GOLDD did not return an identifiability classification for the corresponding model and assumption set.}
	 \label{Softwares_Strike_GOLDD}
\end{table}

\section{Discussion}

Structural identifiability is a fundamental prerequisite for reliable parameter estimation in epidemic modeling, yet it remains underutilized in practice. In this study, we present a user-oriented and reproducible workflow for conducting structural identifiability analysis using the StructuralIdentifiability.jl package and demonstrate its application across a range of commonly used epidemic models. Rather than introducing new identifiability theory, our goal is to illustrate how existing methods can be systematically applied, interpreted, and communicated within applied epidemic modeling contexts. To the best of our knowledge, this is among the first studies to provide a unified, reproducible, and visually interpretable workflow for structural identifiability analysis in epidemic models.\\

Our results show that identifiability depends strongly on model structure, the choice of observed variables, and assumptions regarding initial conditions. In particular, we demonstrate that parameter non-identifiability often arises from structural confounding, such as the coupling between transmission parameters and population size, and that identifiable parameter combinations can still be recovered even when individual parameters are not identifiable.\\

We explored the identifiability of eleven ordinary differential equation-based epidemic models representing a range of infectious disease dynamics. Our aim was to investigate how the progressive addition of epidemiological compartments influences the structural identifiability of model parameters, under scenarios with both known and unknown initial conditions. To further contextualize these findings, we additionally examined identifiability results obtained using the symbolic tools DAISY \cite{bellu2007}, SIAN and STRIKE-GOLDD for selected models. Reporting these results alongside those from \textit{StructuralIdentifiability.jl} highlights areas of agreement and divergence across symbolic approaches, reflecting differences in model structure, observability assumptions, and underlying algorithmic formulations. This comparison is intended to support transparency and cross-validation of identifiability conclusions, rather than to benchmark software performance or computational efficiency. These tools were selected due to their complementary strengths: DAISY is well-established for smaller systems and algebraic clarity, while \texttt{StructuralIdentifiability.jl} offers better scalability and computational efficiency for larger or more complex models. A comparison of their capabilities is provided in Barreiro et al. (2023) \cite{rey2023benchmarking}. From Tables~\ref{Softwares} and~\ref{Softwares_IC}, we observe that the identifiability results are generally consistent between DAISY and \texttt{StructuralIdentifiability.jl}. However, in some cases, DAISY was unable to return the results due to computational limitations or algebraic complexity. In such instances, the \texttt{StructuralIdentifiability.jl} package proved especially useful, successfully producing identifiability results for these models.\\

Importantly, this work does not aim to benchmark identifiability software or to claim computational superiority of any particular tool. Rather, StructuralIdentifiability.jl is used here as a representative implementation to illustrate how global structural identifiability analysis can be practically conducted, interpreted, and communicated in epidemic modeling contexts from a user-oriented perspective.

A predominant practical implication of structural non-identifiability is the need for reparameterization. When individual parameters cannot be uniquely identified, identifiable combinations of parameters can often be estimated instead \cite{saucedo2019computing, wieland2021structural}. While this study focuses on diagnosing identifiability, future work should explore systematic approaches to reparameterization in epidemic models, enabling more robust parameter estimation and model calibration in practice.

One of the key methodological contributions of this work is the practical demonstration of how structural identifiability analysis can be systematically incorporated into the early stages of epidemic model development. The use of the \textit{StructuralIdentifiability.jl} package in Julia provides a powerful yet accessible framework for symbolic analysis, allowing researchers to assess the identifiability of parameters across a diverse set of compartmental models without requiring an extensive background in symbolic computation. From classical SEIR models to more complex systems involving asymptomatic transmission, disease-induced mortality, and vector-borne dynamics, our analyses show how identifiability is influenced by model structure, initial conditions, and the choice (or data availability) of observed outputs. A particularly valuable aspect of our approach is the use of compartmental diagrams annotated with identifiability information, which improves model transparency and facilitates communication among interdisciplinary teams. Furthermore, the application of model reduction techniques using generalized first integrals, particularly in the context of the computationally intensive Ebola transmission model, highlights a promising strategy for extending identifiability assessments to large-scale systems that would otherwise be intractable. Collectively, these methodological advances provide modelers and public health officials with concrete tools to verify the theoretical soundness of their models before engaging in parameter estimation or forecasting, ultimately enhancing the reliability of model-based public health decision-making.\\

Nonetheless, several limitations must be acknowledged. Structural identifiability is a theoretical property that assumes access to noise-free, continuous, and complete data. However, real-world epidemiological data is often sparse, noisy, and subject to reporting delays. In such contexts, practical identifiability analysis - which accounts for finite and noisy data - should be conducted as a complementary step. There are several methodologies that can be implemented for practical identifiability such as the Monte Carlo approach \cite{metropolis1949monte}, the Correlation Matrix \cite{banks2014modeling}, or the Profile Likelihood method \cite{venzon1988method}. Additionally, as the complexity of the model increases, symbolic computation becomes progressively challenging. In our analysis of the Ebola model, we employed a model reduction strategy using generalized first integrals to overcome these computational barriers, an approach that can be extended to other complex systems.\\

In summary, this study demonstrates how symbolic structural identifiability analysis—implemented via the Julia package \textit{StructuralIdentifiability.jl}—can be applied to a broad spectrum of epidemic models, from classical SEIR frameworks to models with increased biological realism and complexity. Compared to our earlier tutorial based on DAISY, this approach enables more efficient analysis of higher-dimensional systems. By systematically exploring how identifiability depends on model structure, observability of outputs, and knowledge of initial conditions, we identify common pitfalls and guide modelers toward more robust formulations. Our annotated flow diagrams and practical model reduction techniques further enhance accessibility and transparency. Future efforts should aim to integrate structural and practical identifiability assessments into unified workflows, thereby improving the reliability and interpretability of models used to inform public health decisions.

\section*{CrediT authorship contribution statement}
YL: Conceptualization, Methodology, Software implementation, Formal analysis, Writing – original draft, Visualization.
OS: Writing – review \& editing.
NT: Supervision, Writing – review \& editing.
GC: Conceptualization, supervision, original draft, Writing – review \& editing.

\section*{Funding}
GC was partially supported by NSF grant DBI 2412115 as part of the US NSF Center for Analysis and Prediction of Pandemic Expansion (APPEX). NT and YL were partially supported by NIH NIGMS 1R01GM152743.

\section*{Declaration of competing interest}
The authors declare that they have no conflict of interest.

\section*{Acknowledgements}
NT and YRL are supported by NIH NIGMS grant no.
1R01GM152743-01. OS is supported by the Simons Foundation Travel Support for Mathematicians 453250. GC is partially supported by NSF grants 2125246 and 2026797.

\section*{Data availability}
All codes used in this study is publicly available at \href{https://github.com/YuganthiLiyanage/Structural-identifiability-of-epidemic-models}{this GitHub repository}.

 \bibliographystyle{elsarticle-num-names} 
 \bibliography{References}

\begin{thebibliography}{48}
\expandafter\ifx\csname natexlab\endcsname\relax\def\natexlab#1{#1}\fi
\providecommand{\url}[1]{\texttt{#1}}
\providecommand{\href}[2]{#2}
\providecommand{\path}[1]{#1}
\providecommand{\DOIprefix}{doi:}
\providecommand{\ArXivprefix}{arXiv:}
\providecommand{\URLprefix}{URL: }
\providecommand{\Pubmedprefix}{pmid:}
\providecommand{\doi}[1]{\href{http://dx.doi.org/#1}{\path{#1}}}
\providecommand{\Pubmed}[1]{\href{pmid:#1}{\path{#1}}}
\providecommand{\bibinfo}[2]{#2}
\ifx\xfnm\relax \def\xfnm[#1]{\unskip,\space#1}\fi
\bibitem[{Brauer(2008)}]{brauer2008}
\bibinfo{author}{F.~Brauer},
\newblock \bibinfo{title}{Compartmental models in epidemiology},
\newblock in: \bibinfo{editor}{F.~Brauer}, \bibinfo{editor}{P.~van~den
  Driessche}, \bibinfo{editor}{J.~Wu} (Eds.), \bibinfo{booktitle}{Mathematical
  Epidemiology}, volume \bibinfo{volume}{1945} of
  \textit{\bibinfo{series}{Lecture Notes in Mathematics}},
  \bibinfo{publisher}{Springer}, \bibinfo{year}{2008}, pp.
  \bibinfo{pages}{19--79}. \DOIprefix\doi{10.1007/978-3-540-78911-6_2}.
\bibitem[{Arino et~al.(2007)Arino, Brauer, van~den Driessche, Watmough, and
  Wu}]{arino2007}
\bibinfo{author}{J.~Arino}, \bibinfo{author}{F.~Brauer},
  \bibinfo{author}{P.~van~den Driessche}, \bibinfo{author}{J.~Watmough},
  \bibinfo{author}{J.~Wu},
\newblock \bibinfo{title}{A final size relation for epidemic models},
\newblock \bibinfo{journal}{Mathematical Biosciences and Engineering}
  \bibinfo{volume}{4} (\bibinfo{year}{2007}) \bibinfo{pages}{159--175}.
  \DOIprefix\doi{10.3934/mbe.2007.4.159}.
\bibitem[{Chowell et~al.(2004)Chowell, Hengartner, Castillo-Chavez, Fenimore,
  and Hyman}]{chowell2004}
\bibinfo{author}{G.~Chowell}, \bibinfo{author}{N.~W. Hengartner},
  \bibinfo{author}{C.~Castillo-Chavez}, \bibinfo{author}{P.~W. Fenimore},
  \bibinfo{author}{J.~M. Hyman},
\newblock \bibinfo{title}{The basic reproductive number of ebola and the
  effects of public health measures: The cases of congo and uganda},
\newblock \bibinfo{journal}{Journal of Theoretical Biology}
  \bibinfo{volume}{229} (\bibinfo{year}{2004}) \bibinfo{pages}{119--126}.
  \DOIprefix\doi{10.1016/j.jtbi.2004.03.006}.
\bibitem[{Anderson et~al.(2004)Anderson, Fraser, Ghani, Donnelly, Riley,
  Ferguson, Leung, Lam, Hedley, Ho et~al.}]{anderson2004}
\bibinfo{author}{R.~M. Anderson}, \bibinfo{author}{C.~Fraser},
  \bibinfo{author}{A.~C. Ghani}, \bibinfo{author}{C.~A. Donnelly},
  \bibinfo{author}{S.~Riley}, \bibinfo{author}{N.~M. Ferguson},
  \bibinfo{author}{G.~M. Leung}, \bibinfo{author}{T.~H. Lam},
  \bibinfo{author}{A.~J. Hedley}, \bibinfo{author}{L.-M. Ho}, et~al.,
\newblock \bibinfo{title}{Epidemiology, transmission dynamics and control of
  sars: The 2002–2003 epidemic},
\newblock \bibinfo{journal}{Philosophical Transactions of the Royal Society of
  London. Series B: Biological Sciences} \bibinfo{volume}{359}
  (\bibinfo{year}{2004}) \bibinfo{pages}{1091--1105}.
  \DOIprefix\doi{10.1098/rstb.2004.1490}.
\bibitem[{Saucedo et~al.(2019)Saucedo, Martcheva, and
  Annor}]{saucedo2019computing}
\bibinfo{author}{O.~Saucedo}, \bibinfo{author}{M.~Martcheva},
  \bibinfo{author}{A.~Annor},
\newblock \bibinfo{title}{Computing human-to-human avian influenza $r_0$ via
  transmission chains and parameter estimation},
\newblock \bibinfo{journal}{Mathematical Biosciences and Engineering}
  \bibinfo{volume}{16} (\bibinfo{year}{2019}) \bibinfo{pages}{3465--3487}.
\bibitem[{Yan and Chowell(2019)}]{yan2019}
\bibinfo{author}{P.~Yan}, \bibinfo{author}{G.~Chowell},
  \bibinfo{title}{Quantitative Methods for Investigating Infectious Disease
  Outbreaks}, \bibinfo{publisher}{Springer}, \bibinfo{year}{2019}.
  \DOIprefix\doi{10.1007/978-3-030-21923-9}.
\bibitem[{Banks and Tran(2009)}]{banks2009}
\bibinfo{author}{H.~T. Banks}, \bibinfo{author}{H.~T. Tran},
  \bibinfo{title}{Mathematical and Experimental Modeling of Physical and
  Biological Processes}, \bibinfo{publisher}{Chapman and Hall/CRC},
  \bibinfo{year}{2009}.
\bibitem[{Tuncer and Le(2018)}]{tuncer2018}
\bibinfo{author}{N.~Tuncer}, \bibinfo{author}{T.~T. Le},
\newblock \bibinfo{title}{Structural and practical identifiability analysis of
  outbreak models},
\newblock \bibinfo{journal}{Mathematical Biosciences} \bibinfo{volume}{299}
  (\bibinfo{year}{2018}) \bibinfo{pages}{1--18}.
  \DOIprefix\doi{10.1016/j.mbs.2018.02.004}.
\bibitem[{Roosa and Chowell(2019)}]{roosa2019}
\bibinfo{author}{K.~Roosa}, \bibinfo{author}{G.~Chowell},
\newblock \bibinfo{title}{Assessing parameter identifiability in compartmental
  dynamic models using a computational approach: Application to infectious
  disease transmission models},
\newblock \bibinfo{journal}{Theoretical Biology and Medical Modelling}
  \bibinfo{volume}{16} (\bibinfo{year}{2019}) \bibinfo{pages}{1}.
  \DOIprefix\doi{10.1186/s12976-018-0097-6}.
\bibitem[{Villaverde et~al.(2016)Villaverde, Barreiro, and
  Papachristodoulou}]{villaverde2016}
\bibinfo{author}{A.~F. Villaverde}, \bibinfo{author}{A.~Barreiro},
  \bibinfo{author}{A.~Papachristodoulou},
\newblock \bibinfo{title}{Structural identifiability of dynamic systems biology
  models},
\newblock \bibinfo{journal}{PLOS Computational Biology} \bibinfo{volume}{12}
  (\bibinfo{year}{2016}) \bibinfo{pages}{e1005153}.
  \DOIprefix\doi{10.1371/journal.pcbi.1005153}.
\bibitem[{Eisenberg et~al.(2013)Eisenberg, Robertson, and Tien}]{eisenberg2013}
\bibinfo{author}{M.~C. Eisenberg}, \bibinfo{author}{S.~L. Robertson},
  \bibinfo{author}{J.~H. Tien},
\newblock \bibinfo{title}{Identifiability and estimation of multiple
  transmission pathways in cholera and waterborne disease},
\newblock \bibinfo{journal}{Journal of Theoretical Biology}
  \bibinfo{volume}{324} (\bibinfo{year}{2013}) \bibinfo{pages}{84--102}.
  \DOIprefix\doi{10.1016/j.jtbi.2012.12.021}.
\bibitem[{Massonis et~al.(2021)Massonis, Banga, and
  Villaverde}]{massonis2021structural}
\bibinfo{author}{G.~Massonis}, \bibinfo{author}{J.~R. Banga},
  \bibinfo{author}{A.~F. Villaverde},
\newblock \bibinfo{title}{Structural identifiability and observability of
  compartmental models of the covid-19 pandemic},
\newblock \bibinfo{journal}{Annual Reviews in Control} \bibinfo{volume}{51}
  (\bibinfo{year}{2021}) \bibinfo{pages}{441--459}.
\bibitem[{Dankwa et~al.(2022)Dankwa, Brouwer, and
  Donnelly}]{dankwa2022structural}
\bibinfo{author}{E.~A. Dankwa}, \bibinfo{author}{A.~F. Brouwer},
  \bibinfo{author}{C.~A. Donnelly},
\newblock \bibinfo{title}{Structural identifiability of compartmental models
  for infectious disease transmission is influenced by data type},
\newblock \bibinfo{journal}{Epidemics} \bibinfo{volume}{41}
  (\bibinfo{year}{2022}) \bibinfo{pages}{100643}.
\bibitem[{Gallo et~al.(2022)Gallo, Frasca, Latora, and Russo}]{gallo2022lack}
\bibinfo{author}{L.~Gallo}, \bibinfo{author}{M.~Frasca},
  \bibinfo{author}{V.~Latora}, \bibinfo{author}{G.~Russo},
\newblock \bibinfo{title}{Lack of practical identifiability may hamper reliable
  predictions in covid-19 epidemic models},
\newblock \bibinfo{journal}{Science Advances} \bibinfo{volume}{8}
  (\bibinfo{year}{2022}) \bibinfo{pages}{eabg5234}.
\bibitem[{Heinrich et~al.(2025)Heinrich, Rosenblatt, Wieland, Stigter, and
  Timmer}]{heinrich2025structural}
\bibinfo{author}{M.~Heinrich}, \bibinfo{author}{M.~Rosenblatt},
  \bibinfo{author}{F.-G. Wieland}, \bibinfo{author}{H.~Stigter},
  \bibinfo{author}{J.~Timmer},
\newblock \bibinfo{title}{On structural and practical identifiability: Current
  status and update of results},
\newblock \bibinfo{journal}{Current Opinion in Systems Biology}
  \bibinfo{volume}{41} (\bibinfo{year}{2025}) \bibinfo{pages}{100546}.
\bibitem[{Rey~Barreiro and Villaverde(2023)}]{rey2023benchmarking}
\bibinfo{author}{X.~Rey~Barreiro}, \bibinfo{author}{A.~F. Villaverde},
\newblock \bibinfo{title}{Benchmarking tools for a priori identifiability
  analysis},
\newblock \bibinfo{journal}{Bioinformatics} \bibinfo{volume}{39}
  (\bibinfo{year}{2023}) \bibinfo{pages}{btad065}.
\bibitem[{Bellu et~al.(2007)Bellu, Saccomani, Audoly, and D'Angiò}]{bellu2007}
\bibinfo{author}{G.~Bellu}, \bibinfo{author}{M.~P. Saccomani},
  \bibinfo{author}{S.~Audoly}, \bibinfo{author}{L.~D'Angiò},
\newblock \bibinfo{title}{Daisy: A new software tool to test global
  identifiability of biological and physiological systems},
\newblock \bibinfo{journal}{Computer Methods and Programs in Biomedicine}
  \bibinfo{volume}{88} (\bibinfo{year}{2007}) \bibinfo{pages}{52--61}.
  \DOIprefix\doi{10.1016/j.cmpb.2007.07.002}.
\bibitem[{Chowell et~al.(2023)Chowell, Dahal, Liyanage, Tariq, and
  Tuncer}]{Chowell2023}
\bibinfo{author}{G.~Chowell}, \bibinfo{author}{S.~Dahal},
  \bibinfo{author}{Y.~R. Liyanage}, \bibinfo{author}{A.~Tariq},
  \bibinfo{author}{N.~Tuncer},
\newblock \bibinfo{title}{Structural identifiability analysis of epidemic
  models based on differential equations: A tutorial-based primer},
\newblock \bibinfo{journal}{Journal of Mathematical Biology}
  \bibinfo{volume}{87} (\bibinfo{year}{2023}) \bibinfo{pages}{79}.
\bibitem[{Dong et~al.(2023)Dong, Goodbrake, Harrington, and
  Pogudin}]{structidjl2023}
\bibinfo{author}{R.~Dong}, \bibinfo{author}{C.~Goodbrake},
  \bibinfo{author}{H.~Harrington}, \bibinfo{author}{G.~Pogudin},
\newblock \bibinfo{title}{Differential elimination for dynamical models via
  projections with applications to structural identifiability},
\newblock \bibinfo{journal}{SIAM Journal on Applied Algebra and Geometry}
  \bibinfo{volume}{7} (\bibinfo{year}{2023}) \bibinfo{pages}{194--235}.
  \DOIprefix\doi{10.1137/22M1469067}.
\bibitem[{Pohjanpalo(1978)}]{pohjanpalo1978}
\bibinfo{author}{H.~Pohjanpalo},
\newblock \bibinfo{title}{System identifiability based on the power series
  expansion of the solution},
\newblock \bibinfo{journal}{Mathematical Biosciences} \bibinfo{volume}{41}
  (\bibinfo{year}{1978}) \bibinfo{pages}{21--33}.
  \DOIprefix\doi{10.1016/0025-5564(78)90063-9}.
\bibitem[{Walter and Lecourtier(1982)}]{walter1982}
\bibinfo{author}{E.~Walter}, \bibinfo{author}{Y.~Lecourtier},
\newblock \bibinfo{title}{Global approaches to identifiability testing for
  linear and nonlinear state space models},
\newblock \bibinfo{journal}{Mathematical Biosciences} \bibinfo{volume}{59}
  (\bibinfo{year}{1982}) \bibinfo{pages}{1--20}.
  \DOIprefix\doi{10.1016/0025-5564(82)90045-0}.
\bibitem[{Vajda et~al.(1989)Vajda, Godfrey, and Rabitz}]{vajda1989}
\bibinfo{author}{S.~Vajda}, \bibinfo{author}{K.~R. Godfrey},
  \bibinfo{author}{H.~Rabitz},
\newblock \bibinfo{title}{Similarity transformation approach to identifiability
  analysis of nonlinear compartmental models},
\newblock \bibinfo{journal}{Mathematical Biosciences} \bibinfo{volume}{93}
  (\bibinfo{year}{1989}) \bibinfo{pages}{217--248}.
  \DOIprefix\doi{10.1016/0025-5564(89)90054-7}.
\bibitem[{Joly-Blanchard and Denis-Vidal(1998)}]{joly1998some}
\bibinfo{author}{G.~Joly-Blanchard}, \bibinfo{author}{L.~Denis-Vidal},
\newblock \bibinfo{title}{Some remarks about an identifiability result of
  nonlinear systems},
\newblock \bibinfo{journal}{Automatica} \bibinfo{volume}{34}
  (\bibinfo{year}{1998}) \bibinfo{pages}{1151--1152}.
\bibitem[{Ljung and Glad(1994)}]{ljung1994}
\bibinfo{author}{L.~Ljung}, \bibinfo{author}{T.~Glad},
\newblock \bibinfo{title}{On global identifiability for arbitrary model
  parametrizations},
\newblock \bibinfo{journal}{Automatica} \bibinfo{volume}{30}
  (\bibinfo{year}{1994}) \bibinfo{pages}{265--276}.
  \DOIprefix\doi{10.1016/0005-1098(94)90029-9}.
\bibitem[{Liyanage et~al.(2025)Liyanage, Chowell, Pogudin, and
  Tuncer}]{phenom2025}
\bibinfo{author}{Y.~R. Liyanage}, \bibinfo{author}{G.~Chowell},
  \bibinfo{author}{G.~Pogudin}, \bibinfo{author}{N.~Tuncer},
\newblock \bibinfo{title}{Structural and practical identifiability of
  phenomenological growth models for epidemic forecasting},
\newblock \bibinfo{journal}{Viruses} \bibinfo{volume}{17}
  (\bibinfo{year}{2025}) \bibinfo{pages}{496}.
\bibitem[{Miao et~al.(2011)Miao, Xia, Perelson, and
  Wu}]{miao2011identifiability}
\bibinfo{author}{H.~Miao}, \bibinfo{author}{X.~Xia}, \bibinfo{author}{A.~S.
  Perelson}, \bibinfo{author}{H.~Wu},
\newblock \bibinfo{title}{On identifiability of nonlinear ode models and
  applications in viral dynamics},
\newblock \bibinfo{journal}{SIAM review} \bibinfo{volume}{53}
  (\bibinfo{year}{2011}) \bibinfo{pages}{3--39}.
\bibitem[{Saucedo et~al.(2024)Saucedo, Laubmeier, Tang, Levy, Asik, Pollington,
  and Prosper}]{Saucedo2024Indent}
\bibinfo{author}{O.~Saucedo}, \bibinfo{author}{A.~Laubmeier},
  \bibinfo{author}{T.~Tang}, \bibinfo{author}{B.~Levy},
  \bibinfo{author}{L.~Asik}, \bibinfo{author}{T.~Pollington},
  \bibinfo{author}{O.~Prosper},
\newblock \bibinfo{title}{Comparative analysis of practical identifiability
  methods for an {SEIR} model},
\newblock \bibinfo{journal}{AIMS Mathematics} \bibinfo{volume}{9}
  (\bibinfo{year}{2024}) \bibinfo{pages}{24722--24761}.
\bibitem[{Ligon et~al.(2018)Ligon, Fr{\"o}hlich, Chi{\c{s}}, Banga,
  Balsa-Canto, and Hasenauer}]{ligon2018genssi}
\bibinfo{author}{T.~S. Ligon}, \bibinfo{author}{F.~Fr{\"o}hlich},
  \bibinfo{author}{O.~T. Chi{\c{s}}}, \bibinfo{author}{J.~R. Banga},
  \bibinfo{author}{E.~Balsa-Canto}, \bibinfo{author}{J.~Hasenauer},
\newblock \bibinfo{title}{Genssi 2.0: multi-experiment structural
  identifiability analysis of sbml models},
\newblock \bibinfo{journal}{Bioinformatics} \bibinfo{volume}{34}
  (\bibinfo{year}{2018}) \bibinfo{pages}{1421--1423}.
\bibitem[{D{\'\i}az-Seoane et~al.(2023)D{\'\i}az-Seoane, Rey~Barreiro, and
  Villaverde}]{diaz2023strike}
\bibinfo{author}{S.~D{\'\i}az-Seoane}, \bibinfo{author}{X.~Rey~Barreiro},
  \bibinfo{author}{A.~F. Villaverde},
\newblock \bibinfo{title}{Strike-goldd 4.0: user-friendly, efficient analysis
  of structural identifiability and observability},
\newblock \bibinfo{journal}{Bioinformatics} \bibinfo{volume}{39}
  (\bibinfo{year}{2023}) \bibinfo{pages}{btac748}.
\bibitem[{Anstett-Collin et~al.(2020)Anstett-Collin, Denis-Vidal, and
  Mill{\'e}rioux}]{anstett2020priori}
\bibinfo{author}{F.~Anstett-Collin}, \bibinfo{author}{L.~Denis-Vidal},
  \bibinfo{author}{G.~Mill{\'e}rioux},
\newblock \bibinfo{title}{A priori identifiability: An overview on definitions
  and approaches},
\newblock \bibinfo{journal}{Annual Reviews in Control} \bibinfo{volume}{50}
  (\bibinfo{year}{2020}) \bibinfo{pages}{139--149}.
\bibitem[{Castro and De~Boer(2020)}]{castro2020testing}
\bibinfo{author}{M.~Castro}, \bibinfo{author}{R.~J. De~Boer},
\newblock \bibinfo{title}{Testing structural identifiability by a simple
  scaling method},
\newblock \bibinfo{journal}{PLOS Computational Biology} \bibinfo{volume}{16}
  (\bibinfo{year}{2020}) \bibinfo{pages}{e1008248}.
\bibitem[{Saccomani et~al.(2003)Saccomani, Audoly, and
  D'Angi{\`o}}]{saccomani2003parameter}
\bibinfo{author}{M.~P. Saccomani}, \bibinfo{author}{S.~Audoly},
  \bibinfo{author}{L.~D'Angi{\`o}},
\newblock \bibinfo{title}{Parameter identifiability of nonlinear systems: the
  role of initial conditions},
\newblock \bibinfo{journal}{Automatica} \bibinfo{volume}{39}
  (\bibinfo{year}{2003}) \bibinfo{pages}{619--632}.
\bibitem[{Guillaume et~al.(2019)Guillaume, Jakeman, Marsili-Libelli, Asher,
  Brunner, Croke, Hill, Jakeman, Keesman, Razavi
  et~al.}]{guillaume2019introductory}
\bibinfo{author}{J.~H.~A. Guillaume}, \bibinfo{author}{J.~D. Jakeman},
  \bibinfo{author}{S.~Marsili-Libelli}, \bibinfo{author}{M.~Asher},
  \bibinfo{author}{P.~Brunner}, \bibinfo{author}{B.~Croke},
  \bibinfo{author}{M.~C. Hill}, \bibinfo{author}{A.~J. Jakeman},
  \bibinfo{author}{K.~J. Keesman}, \bibinfo{author}{S.~Razavi}, et~al.,
\newblock \bibinfo{title}{Introductory overview of identifiability analysis: A
  guide to evaluating whether you have the right type of data for your modeling
  purpose},
\newblock \bibinfo{journal}{Environmental Modelling \& Software}
  \bibinfo{volume}{119} (\bibinfo{year}{2019}) \bibinfo{pages}{418--432}.
\bibitem[{Cobelli and Distefano(1980)}]{cobelli1980parameter}
\bibinfo{author}{C.~Cobelli}, \bibinfo{author}{J.~J. Distefano},
\newblock \bibinfo{title}{Parameter and structural identifiability concepts and
  ambiguities: A critical review and analysis},
\newblock \bibinfo{journal}{American Journal of Physiology-Regulatory,
  Integrative and Comparative Physiology} \bibinfo{volume}{239}
  (\bibinfo{year}{1980}) \bibinfo{pages}{R7--R24}.
\bibitem[{Bellman and {\AA}str{\"o}m(1970)}]{bellman1970structural}
\bibinfo{author}{R.~Bellman}, \bibinfo{author}{K.~J. {\AA}str{\"o}m},
\newblock \bibinfo{title}{On structural identifiability},
\newblock \bibinfo{journal}{Mathematical Biosciences} \bibinfo{volume}{7}
  (\bibinfo{year}{1970}) \bibinfo{pages}{329--339}.
\bibitem[{Distefano and Cobelli(1980)}]{distefano1980parameter}
\bibinfo{author}{J.~Distefano}, \bibinfo{author}{C.~Cobelli},
\newblock \bibinfo{title}{On parameter and structural identifiability:
  Nonunique observability/reconstructibility for identifiable systems, other
  ambiguities, and new definitions},
\newblock \bibinfo{journal}{IEEE Transactions on Automatic Control}
  \bibinfo{volume}{25} (\bibinfo{year}{1980}) \bibinfo{pages}{830--833}.
  \DOIprefix\doi{10.1109/TAC.1980.1102363}.
\bibitem[{Ollivier(1990)}]{Ollivier1990}
\bibinfo{author}{M.~Ollivier}, \bibinfo{title}{Le Problème de
  l'Identifiabilité Structurelle Globale: Approche Théorique, Méthodes
  Effectives et Bornes de Complexité}, \bibinfo{type}{Ph.d. thesis}, École
  Polytechnique, \bibinfo{address}{Paris}, \bibinfo{year}{1990}.
\bibitem[{Hong et~al.(2020)Hong, Ovchinnikov, Pogudin, and Yap}]{Hong2020}
\bibinfo{author}{H.~Hong}, \bibinfo{author}{A.~Ovchinnikov},
  \bibinfo{author}{G.~Pogudin}, \bibinfo{author}{C.~Yap},
\newblock \bibinfo{title}{Global identifiability of differential models},
\newblock \bibinfo{journal}{Communications on Pure and Applied Mathematics}
  \bibinfo{volume}{73} (\bibinfo{year}{2020}) \bibinfo{pages}{1831--1879}.
  \DOIprefix\doi{10.1002/cpa.21921}.
\bibitem[{Pogudin(2024)}]{Pogudin2024}
\bibinfo{author}{G.~Pogudin}, \bibinfo{title}{Generalized first integrals and
  structural identifiability}, \bibinfo{year}{2024}. \URLprefix
  \url{https://github.com/SciML/StructuralIdentifiability.jl/issues/63},
  \bibinfo{note}{accessed: 2025-10-11}.
\bibitem[{Ballif et~al.(2024)Ballif, Clément, and Yvinec}]{Ballif2024}
\bibinfo{author}{G.~Ballif}, \bibinfo{author}{F.~Clément},
  \bibinfo{author}{R.~Yvinec},
\newblock \bibinfo{title}{Nonlinear compartmental modeling to monitor ovarian
  follicle population dynamics on the whole lifespan},
\newblock \bibinfo{journal}{Journal of Mathematical Biology}
  \bibinfo{volume}{89} (\bibinfo{year}{2024}) \bibinfo{pages}{9}.
  \DOIprefix\doi{10.1007/s00285-024-02108-6}.
\bibitem[{Meshkat et~al.(2014)Meshkat, Kuo, and DiStefano}]{combos}
\bibinfo{author}{N.~Meshkat}, \bibinfo{author}{C.~Kuo}, \bibinfo{author}{J.~r.
  DiStefano},
\newblock \bibinfo{title}{On finding and using identifiable parameter
  combinations in nonlinear dynamic systems biology models and {COMBOS}: {A}
  novel web implementation},
\newblock \bibinfo{journal}{PLos One} \bibinfo{volume}{9}
  (\bibinfo{year}{2014}).
\bibitem[{Hong et~al.(2019)Hong, Ovchinnikov, Pogudin, and Yap}]{SIAN}
\bibinfo{author}{H.~Hong}, \bibinfo{author}{A.~Ovchinnikov},
  \bibinfo{author}{G.~Pogudin}, \bibinfo{author}{C.~Yap},
\newblock \bibinfo{title}{Sian: software for structural identifiability
  analysis of ode models},
\newblock \bibinfo{journal}{Bioinformatics} \bibinfo{volume}{35}
  (\bibinfo{year}{2019}) \bibinfo{pages}{2873--2874}. \URLprefix
  \url{https://doi.org/10.1093/bioinformatics/bty1069}.
  \DOIprefix\doi{10.1093/bioinformatics/bty1069}.
  \href{http://arxiv.org/abs/https://academic.oup.com/bioinformatics/article-pdf/35/16/2873/50719225/bty1069.pdf}{{\tt
  arXiv:https://academic.oup.com/bioinformatics/article-pdf/35/16/2873/50719225/bty1069.pdf}}.
\bibitem[{Villaverde et~al.(2016)Villaverde, Barreiro, and
  Papachristodoulou}]{Strike}
\bibinfo{author}{A.~F. Villaverde}, \bibinfo{author}{A.~Barreiro},
  \bibinfo{author}{A.~Papachristodoulou},
\newblock \bibinfo{title}{Structural identifiability of dynamic systems biology
  models},
\newblock \bibinfo{journal}{PLoS computational biology} \bibinfo{volume}{12}
  (\bibinfo{year}{2016}).
\bibitem[{Villaverde et~al.(2019)Villaverde, Tsiantis, and Banga}]{Strike2}
\bibinfo{author}{A.~F. Villaverde}, \bibinfo{author}{N.~Tsiantis},
  \bibinfo{author}{J.~R. Banga},
\newblock \bibinfo{title}{Full observability and estimation of unknown inputs,
  states and parameters of nonlinear biological models},
\newblock \bibinfo{journal}{Journal of The Royal Society Interface}
  \bibinfo{volume}{16} (\bibinfo{year}{2019}) \bibinfo{pages}{20190043}.
  \URLprefix \url{https://doi.org/10.1098/rsif.2019.0043}.
  \DOIprefix\doi{10.1098/rsif.2019.0043}.
\bibitem[{Wieland et~al.(2021)Wieland, Hauber, Rosenblatt, T{\"o}nsing, and
  Timmer}]{wieland2021structural}
\bibinfo{author}{F.-G. Wieland}, \bibinfo{author}{A.~L. Hauber},
  \bibinfo{author}{M.~Rosenblatt}, \bibinfo{author}{C.~T{\"o}nsing},
  \bibinfo{author}{J.~Timmer},
\newblock \bibinfo{title}{On structural and practical identifiability},
\newblock \bibinfo{journal}{Current opinion in systems biology}
  \bibinfo{volume}{25} (\bibinfo{year}{2021}) \bibinfo{pages}{60--69}.
\bibitem[{Metropolis and Ulam(1949)}]{metropolis1949monte}
\bibinfo{author}{N.~Metropolis}, \bibinfo{author}{S.~Ulam},
\newblock \bibinfo{title}{The monte carlo method},
\newblock \bibinfo{journal}{Journal of the American Statistical Association}
  \bibinfo{volume}{44} (\bibinfo{year}{1949}) \bibinfo{pages}{335--341}.
\bibitem[{Banks et~al.(2014)Banks, Hu, and Thompson}]{banks2014modeling}
\bibinfo{author}{H.~T. Banks}, \bibinfo{author}{S.~Hu}, \bibinfo{author}{W.~C.
  Thompson}, \bibinfo{title}{Modeling and Inverse Problems in the Presence of
  Uncertainty}, \bibinfo{publisher}{CRC Press}, \bibinfo{year}{2014}.
\bibitem[{Venzon and Moolgavkar(1988)}]{venzon1988method}
\bibinfo{author}{D.~J. Venzon}, \bibinfo{author}{S.~H. Moolgavkar},
\newblock \bibinfo{title}{A method for computing profile-likelihood-based
  confidence intervals},
\newblock \bibinfo{journal}{Journal of the Royal Statistical Society: Series C
  (Applied Statistics)} \bibinfo{volume}{37} (\bibinfo{year}{1988})
  \bibinfo{pages}{87--94}.

\end{thebibliography}







\end{document}